\documentclass[sn-mathphys,iicol]{sn-jnl}

\usepackage{graphicx}
\usepackage{dcolumn}
\usepackage{bm}
\usepackage{mathtools}

\graphicspath{{figures/pdf/}}

\begin{document}

\title{Discovering Conservation Laws using Optimal Transport and Manifold Learning}


\author*[1,2]{\fnm{Peter Y.} \sur{Lu}}\email{lup@uchicago.edu}
\author[3]{\fnm{Rumen} \sur{Dangovski}}
\author[2]{\fnm{Marin} \sur{Solja\v{c}i\'{c}}}

\affil[1]{\orgdiv{Data Science Institute}, \orgname{University of Chicago}, \orgaddress{\city{Chicago}, \state{IL} \postcode{60637}, \country{USA}}}

\affil[2]{\orgdiv{Department of Physics}, \orgname{Massachusetts Institute of Technology}, \orgaddress{\city{Cambridge}, \state{MA} \postcode{02139}, \country{USA}}}

\affil[3]{\orgdiv{Department of Electrical Engineering and Computer Science}, \orgname{Massachusetts Institute of Technology}, \orgaddress{\city{Cambridge}, \state{MA} \postcode{02139}, \country{USA}}}


\abstract{Conservation laws are key theoretical and practical tools for understanding, characterizing, and modeling nonlinear dynamical systems. However, for many complex systems, the corresponding conserved quantities are difficult to identify, making it hard to analyze their dynamics and build stable predictive models. Current approaches for discovering conservation laws often depend on detailed dynamical information or rely on black box parametric deep learning methods. We instead reformulate this task as a manifold learning problem and propose a non-parametric approach for discovering conserved quantities. We test this new approach on a variety of physical systems and demonstrate that our method is able to both identify the number of conserved quantities and extract their values. Using tools from optimal transport theory and manifold learning, our proposed method provides a direct geometric approach to identifying conservation laws that is both robust and interpretable without requiring an explicit model of the system nor accurate time information.}




\maketitle

\section{Introduction}

Conservation laws are powerful constraints on the dynamics of many physical systems in nature, and the corresponding conserved quantities are essential features for characterizing the behavior of these systems. Through Noether's theorem, conservation laws are closely tied with the symmetries of a physical system and play a key role in our understanding of physics. Conservation laws also help stabilize and enhance the performance of predictive models for complex nonlinear dynamics, e.g.\ symplectic integrators for Hamiltonian systems \cite{Hairer2006} and pressure projection for incompressible fluid flow \cite{GUERMOND20066011}. In fact, for chaotic dynamical systems, conserved quantities are often the only features of the system state that can be reliably known far into the future. Discovering conservation laws helps us characterize the long term behavior of complex dynamical systems and understand the underlying physics.

While the conservation laws of many physical systems are well-known and often derived from known symmetries, there are still many instances where it is difficult to even determine the number of conservation laws, let alone explicitly extract the conserved quantities. As a historical example, consider the Korteweg--De Vries (KdV) equation modeling shallow water waves. The KdV equation, despite its apparent complexity, has infinitely many conserved quantities \cite{doi:10.1063/1.1664701} and is, in fact, fully solvable via an inverse scattering transform \cite{PhysRevLett.19.1095}---a discovery made after significant theoretical and computational effort. Developing better general methods for identifying conserved quantities will allow us to improve our understanding of new or understudied physical systems and build more efficient and stable predictive models.

In real-world applications, an accurate model for the underlying physical system is often unavailable, forcing us to identify conservation laws using only sample trajectories of the system dynamics. One broad approach is to use modern data-driven methods based on the Koopman operator formulation of dynamical systems, which lifts the dynamics into an infinite dimensional operator space \cite{Mauroy2020}. In the Koopman formalism, conserved quantities are just one type of Koopman eigenfunction with eigenvalue zero. Thus, one approach is to first apply a system identification method, such as dynamic mode decomposition \cite{schmid_2010,Williams2015ADA}, sparse identification with a library of basis functions \cite{Brunton3932}, or even deep learning-based approaches \cite{lusch2018deep,Champion22445,Lu2022}, to model the system dynamics and then set up and solve the Koopman eigenvalue problem. Alternatively, previous work has also proposed directly setting up the eigenvalue problem by estimating time derivatives from data and then fitting the conserved quantities using a library of possible terms \cite{8618963} or a neural network \cite{https://doi.org/10.48550/arxiv.2203.12610}. These methods can work quite well but require that the measured trajectories have sufficiently low noise and high time resolution in order to accurately estimate time derivatives.

Constructing a model for a dynamical system provides much more information than just the conservation laws. In fact, even estimating time derivatives is usually not necessary if we are only interested in identifying conserved quantities. In this work, we will instead focus on an alternative approach that does not require an explicit model or detailed time information but rather takes advantage of the geometric constraints imposed by conservation laws. Specifically, the presence of conservation laws restricts each trajectory in phase space to lie solely on a lower dimensional isosurface of the conserved quantities. The dimensionality of these isosurfaces can provide information about the number of conserved quantities or constraints \cite{PhysRevLett.126.180604}. Furthermore, since each isosurface corresponds to a particular set of conserved quantities, the variations in shape of the isosurfaces directly correspond to variations in the conserved quantities. In other words, we can identify and extract conserved quantities by examining the varying shapes of the isosurfaces sampled by the trajectories.

In contrast with recent work using black box deep learning methods to fit conserved quantities that are consistent with the sampled isosurfaces \cite{PhysRevResearch.2.033499,PhysRevResearch.3.L042035}, we propose and demonstrate a non-parametric manifold learning approach (Fig.~\ref{fig:method}) that directly characterizes the variations in the sampled isosurfaces, producing an embedding of the space of conserved quantities. Our method first uses the Wasserstein metric from optimal transport \cite{Villani2009} to compute distances in shape space between pairs of sampled isosurfaces and then extracts a low dimensional embedding for the manifold of isosurfaces using diffusion maps \cite{6789755,Coifman2005}. Each point in this embedding corresponds to a distinct isosurface and therefore to a distinct set of conserved quantities, i.e.\ the embedding explicitly parameterizes the space of varying conserved quantities. Related methods have been recently suggested for characterizing molecular conformations using the 1-Wasserstein distance together with diffusion maps \cite{9098723}, performing system identification by comparing invariant measures using the 2-Wasserstein distance \cite{https://doi.org/10.48550/arxiv.2104.15138}, and reconstructing normal forms using diffusion maps \cite{doi:10.1073/pnas.1620045114}. Recent theoretical work has also formalized the idea of using alternative non-Euclidean norms, like the Wasserstein distance, in spectral embedding methods such as diffusion maps \cite{Kileel2021}.

We provide an analytic analysis of our approach for a simple harmonic oscillator system and numerically test our method on several physical systems: the single and double pendulum, planar gravitational dynamics, the KdV equation for shallow water waves, and a nonlinear reaction--diffusion equation that generates an oscillating Turing pattern. We also demonstrate the robustness of our approach to noise in the measured trajectories, missing information in the form of a partially observed phase space, as well as approximate conservation laws (additional experiments in Appendix \ref{apx:robust} and Appendix \ref{apx:langevinsho}). In our comparison tests (Appendix \ref{apx:comparison}), our approach outperforms prior deep learning-based direct fitting methods, all while being an order of magnitude faster. We also provide an easy-to-use codebase, which parallelizes across multiple GPUs, to make an efficient implementation of our method as accessible as possible.\footnote{\url{https://github.com/peterparity/conservation-laws-manifold-learning}}

\section{Proposed Manifold Learning Approach}\label{sec:approach}

\begin{figure*}
    \centering
    \includegraphics[scale=0.88]{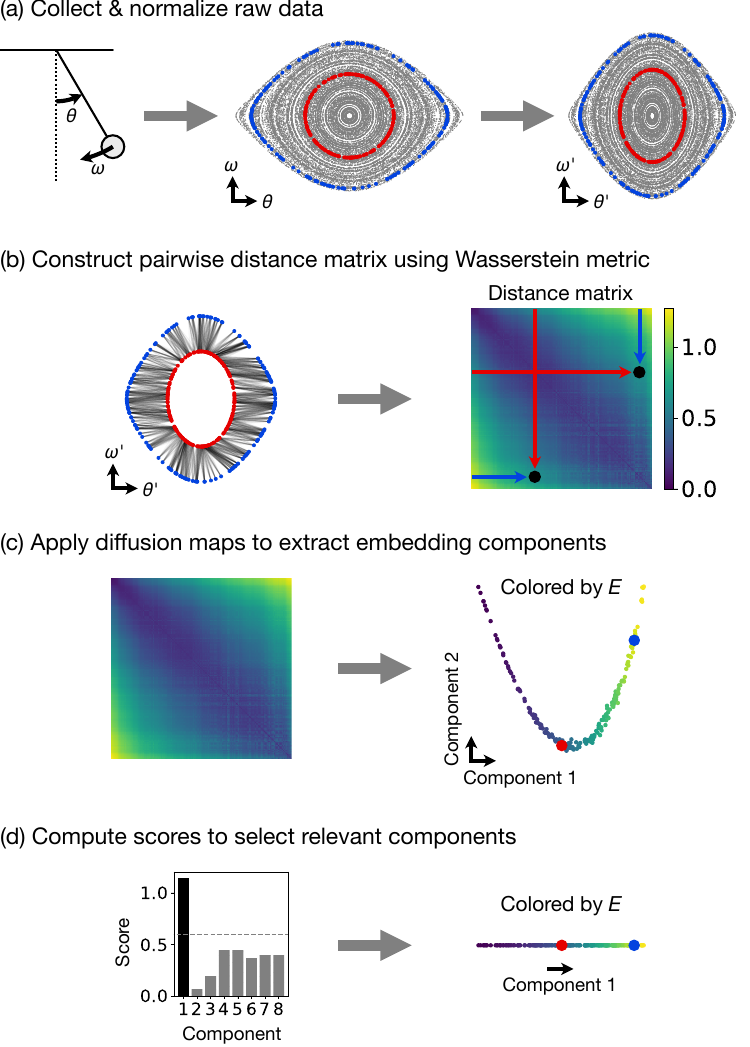}
    \caption{\textbf{Proposed non-parametric method for discovering conservation laws illustrated using a simple pendulum example.} (a) First, we collect and normalize the trajectory data from the dynamical system. Two example trajectories are highlighted in red and blue. (b) Then, we use the Wasserstein metric from optimal transport to compute the distance between each pair of trajectories and construct a distance matrix. For the two example trajectories, the optimal transport plan is shown as lines connecting pairs of points. The constructed distance matrix is plotted with color representing the computed Wasserstein distance between each pair of trajectories. The computed distance between the two example trajectories is marked (black dots) on the distance matrix plot. (c) An embedding of the shape space manifold $\mathcal C$ is extracted from the distance matrix using diffusion maps. The embedding plot is colored by the conserved energy of the pendulum $E$. The points corresponding to the two example trajectories are marked in red and blue. (d) Finally, a heuristic score (Appendix \ref{apx:score}) is used to select relevant components. In this case, only component 1 is relevant, corresponding to a single conserved quantity---the energy $E$. Again, the embedding plot is colored by $E$, and the two example trajectories are marked in red and blue.}
    \label{fig:method}
\end{figure*}

Our proposed approach uses manifold learning to identify and embed the manifold of phase space isosurfaces sampled by the trajectories of a dynamical system. In particular, we compute a diffusion map over a set of trajectories, each of which samples a particular phase space isosurface (Fig.~\ref{fig:method}a). The pairwise distances between these trajectories are given by the 2-Wasserstein distance (Fig.~\ref{fig:method}b), providing the metric structure necessary for applying diffusion maps (Fig.~\ref{fig:method}c). The manifold embedding extracted by the diffusion map corresponds directly to the space of conserved quantities (Fig.~\ref{fig:method}d). Note that this type of analysis does not require knowledge of the equations of motion (Eq.~\ref{eq:system}) and makes no direct reference to time.

\subsection{Dynamical Systems}
Consider a dynamical system with states $\mathbf{x} \in \mathcal M$ that live in a $d$-dimensional phase space $\mathcal M$ and evolve in time according to a system of first order ODEs
\begin{equation}
    \frac{\mathrm{d}\mathbf{x}}{\mathrm{d}t} = \mathbf{F}(\mathbf{x})
    \label{eq:system}
\end{equation}
with $n$ conserved quantities $G_1(\mathbf{x}),\ldots,G_n(\mathbf{x})$.

\subsubsection{Conserved Quantities and Phase Space Isosurfaces}
A conserved quantity $G_i(\mathbf{x})$ is a function of the system state $\mathbf{x}$ that does not change over time, i.e.
\begin{equation}
    \frac{\mathrm{d}G_i(\mathbf{x}(t))}{\mathrm{d}t} = 0,
\end{equation}
but may vary across different initial conditions. As a result, along a particular trajectory $\mathbf{x}(t)$, the $n$ conserved quantities form a set of constraints
\begin{equation}
    G_i(\mathbf{x}) = c_i,\quad i \in \{1,2,\ldots,n\}
    \label{eq:constraint}
\end{equation}
which depend on the values of the conserved quantities $\mathbf{c} = \{c_1,c_2,\ldots,c_n\}$. This set of constraint equations restricts the trajectory to lie in a phase space isosurface $\mathcal X_{\mathbf{c}} \subseteq \mathcal M$ with dimension $d-n$. In fact, if any point of a trajectory lies on the isosurface $\mathcal X_{\mathbf{c}}$, then all other points from the trajectory will lie on the same isosurface.

By studying the variations in shape of these isosurfaces, we are able to directly characterize the space of conserved quantities. In particular, consider the manifold $\mathcal C$ formed by the isosurfaces $\mathcal X_{\mathbf{c}}$ in shape space.\footnote{We are using the term ``manifold'' here rather loosely. While $\mathcal C$ may be a true manifold in many cases, it is also possible for $\mathcal C$ to have non-manifold structure (e.g.\ see our double pendulum experiment in Sec.~\ref{sec:double}).} This manifold $\mathcal C$ is parameterized by the conserved quantities $\mathbf{c}$. Therefore, by analyzing $\mathcal C$ using manifold learning, we can extract the conservation laws of the underlying dynamical system.

\subsubsection{Ergodicity and Physical Measures}\label{sec:ergodicity}

To uniquely identify the isosurface associated with each trajectory, we must make several additional assumptions that will allow us to treat the set of points making up each trajectory as samples from an ergodic invariant measure on the corresponding isosurface. Specifically, we assume that, for each trajectory $\mathbf{x}(t)$ with conserved quantities $\mathbf{c}$, the dynamical system (Eq.~\ref{eq:system}) admits a physical measure \cite{medio_lines_2001} that is ergodic on the isosurface $\mathcal X_{\mathbf{c}}$ and is defined by
\begin{equation}\label{eq:measure}
    \mu_{\mathbf{c}} = \lim_{T\to\infty}\frac{1}{T}\int_0^T\delta_{\mathbf{x}(t)}\,\mathrm{d}t,
\end{equation}
where $\delta_{\mathbf{x}(t)}$ is the Dirac measure centered on $\mathbf{x}(t)$. This ensures that trajectories with the same conserved quantities will sample the same distribution on the same isosurface, allowing us to use the distribution sampled by each trajectory as a proxy for the corresponding isosurface. For more details about this assumption, see Appendix \ref{apx:measure}.

In practice, the sampled distribution may be lower dimensional than the corresponding isosurface if some of the conserved quantities do not vary in the dataset and instead correspond to fixed constraints, or if the dynamical system is dissipative. In the former case, this does not affect our ability to uniquely identify a distribution with an isosurface and its corresponding set of conserved quantities, meaning that we are able to apply this approach even if the provided phase space is much larger than the intrinsic phase space of the dynamical system. In the latter case, the dissipative nature of the system may cause information about conservation laws relevant during the transient portion of the dynamics to be lost, but we are still able to use our approach to identify conserved quantities relevant for the long term behavior of the system.

\subsection{Wasserstein Metric}

To analyze the isosurface shape space manifold $\mathcal C$---i.e.\ the manifold of conserved quantities---using manifold learning methods, we need to place some structure on the points $\mathcal X_{\mathbf{c}} \in \mathcal C$. Having associated each isosurface $\mathcal X_{\mathbf{c}}$ with a corresponding distribution defined by an ergodic physical measure $\mu_{\mathbf{c}}$, we choose to lift the Euclidean metric on the phase space into the space of distributions using the 2-Wasserstein metric from optimal transport
\begin{equation}
    W_2(\mu_{\mathbf{c}}, \mu_{\mathbf{c}'}) = \left(\inf_{\pi \in \Pi(\mu_{\mathbf{c}}, \mu_{\mathbf{c}'})}\int c(\mathbf{x},\mathbf{y})\,\mathrm{d}\pi(\mathbf{x},\mathbf{y})\right)^{1/2},
\end{equation}
where the cost function $c(\mathbf{x},\mathbf{y}) = \lVert\mathbf{x}-\mathbf{y}\rVert^2$ is the squared Euclidean distance, and $\pi \in \Pi(\mu_{\mathbf{c}}, \mu_{\mathbf{c}'})$ is a valid transport map between $\mu_{\mathbf{c}}$ and $\mu_{\mathbf{c}'}$ \cite{Villani2009}.

For discrete samples, the 2-Wasserstein distance between two sets of sample points $\{\mathbf{x}_1,\mathbf{x}_2,\ldots,\mathbf{x}_S\}$ and $\{\mathbf{y}_1,\mathbf{y}_2,\ldots,\mathbf{y}_S\}$ is defined as
\begin{equation}\label{eq:discreteW2}
    W_2 = \bigg(\min_{T} \sum_{i,j}T_{ij}C_{ij}\bigg)^{1/2},
\end{equation}
where the cost matrix $C_{ij} = \lVert\mathbf{x}_i-\mathbf{y}_j\rVert^2$, and the transport matrix $T$ is subject to the constraints
\begin{align}
\begin{split}
    & T_{ij}\ge 0,\quad \forall i,j\\
    & \sum_j T_{ij} = 1\\
    & \sum_i T_{ij} = 1.
\end{split}
\end{align}
To efficiently compute an entropy regularized form of this optimization problem, we use the Sinkhorn algorithm \cite{NIPS2013_af21d0c9} and estimate the Wasserstein distance as a debiased Sinkhorn divergence \cite{pmlr-v119-janati20a}.

One important subtlety in this construction is the choice of the ground metric for optimal transport. As previously mentioned, we use a Euclidean metric on the phase space, which implicitly imposes a choice of units to make the phase space dimensionless. In fact, there is no canonical choice for the ground metric, and different choices result in different Wasserstein metrics on the shape space. While, in theory, information about all conserved quantities will be embedded in the resulting distance matrix regardless of the choice of metric, the metric ultimately determines how easy it is to access this information. For example, when multiple conserved quantities are present, the relative effect of each conserved quantity on the computed Wasserstein distances will determine how prominent each conserved quantity is and how easily it is identified using manifold learning. To partially mitigate this issue and improve consistency, we normalize each component of our data to have a maximum absolute value of 1 before computing the pairwise Wasserstein distances.

Finally, using the Wasserstein distance provides our approach with a tremendous amount of robustness (Appendix \ref{apx:robust}), but also makes it susceptible to certain kinds of sampling inhomogeneity. See Appendix \ref{apx:samplinginhom} for a more detailed discussion of this trade off.

\subsection{Diffusion Maps}

Using the structure provided by the Wasserstein metric, we then use diffusion maps to generate an embedding for $\mathcal C$. The diffusion map manifold learning method uses a spectral embedding algorithm applied to an affinity matrix to construct a low dimensional embedding of the data manifold \cite{6789755,Coifman2005}. Using the pairwise Wasserstein distances $W_2(\mu_i,\mu_j)$ computed from discrete samples provided by the trajectory data (Eq.~\ref{eq:discreteW2}), we first construct a kernel matrix using a Gaussian kernel
\begin{equation}\label{eq:kernel}
    K_{ij} = \exp(-W_2(\mu_i,\mu_j)^2/\epsilon)
\end{equation}
and then scale it to form an affinity matrix for our spectral embedding
\begin{equation}\label{eq:affinity}
    M_{ij} = K_{ij}/(D_iD_j)^\alpha,
\end{equation}
where $D_i = \sum_k K_{ik}$, and $\alpha$ is a hyperparameter. The spectral embedding algorithm then takes this affinity matrix and constructs a normalized graph Laplacian
\begin{equation}
    L_{ij} = I_{ij} - M_{ij}/\sum_k M_{ik},
\end{equation}
where $I$ is the identity matrix. The eigenvectors $\mathbf{v}_i$ corresponding to the smallest eigenvalues $\lambda_i \ge 0$ (excluding $\lambda_0 = 0$) of the Laplacian then provide an approximate low dimensional embedding of the manifold of conserved quantities $\mathcal C$. In our experiments, we set $\alpha=1$ so that the Laplacian computed by the spectral embedding algorithm approximates the Laplace--Beltrami operator \cite{6789755}.

To estimate the dimensionality of $\mathcal C$ and to choose which eigenvectors $\mathbf{v}_i$ to include in our embedding, we use a heuristic score that combines a measure of relevance, given by a length scale computed from the Laplacian eigenvalues, with a previously suggested measure of ``unpredictability'' for minimizing redundancy \cite{pfau2018minimally} (alternative approaches also exist \cite{DSILVA2018759,vonLindheim2018}). To construct our embedding, we only include the Laplacian eigenvectors with score above a chosen cutoff value and discard the rest as either noise or redundant embedding components. To determine the cutoff, we perform a sweep of the cutoff value looking for robust ranges and find that a cutoff of 0.6 works well across all of our experiments, which consist of a wide variety of datasets and dynamical systems. See Appendix \ref{apx:score} for more details.

\section{Analytic Result for the Simple Harmonic Oscillator}\label{sec:analytic}

In the case of a simple harmonic oscillator (SHO) without measurement noise and in the infinite sample limit, we are able to explicitly derive an analytic result for our proposed procedure. We first compute the pairwise distances provided by the Wasserstein metric and then derive the embedding produced by a diffusion map, which corresponds to the conserved energy of the SHO.

\subsection{Wasserstein Metric: Constructing the Isosurface Shape Space}

Consider a SHO with Hamiltonian
\begin{equation}
	H(q,p) = \frac{1}{2m}p^2 + \frac{1}{2}m\omega^2q^2
\end{equation}
given in terms of position $q$ and momentum $p$. The SHO energy isosurfaces $E = H(q,p)$ form concentric ellipses in a 2D phase space. Choosing units such that $m = 1$ and $\omega = 1$, we obtain concentric circles with uniformly distributed samples (assuming a uniform sampling in time). The 2-Wasserstein distance between a pair of uniformly distributed circular isosurfaces is simply given by the difference in radii $\lvert r_1 - r_2\rvert$. This is because, due to the rotational symmetry of the two distributions, the optimal transport plan for an isotropic cost function is to simply move each point on isosurface 1 radially outward (or inward) to the point on isosurface 2 with the same angle $\theta$.

This result does not meaningfully change with a different choice of units, which is equivalent to rescaling the phase space coordinates $q, p$. If we rescale $q, p$ by factors $k_q, k_p$, our cost function simply becomes
\begin{align}
\begin{split}
	c(\theta_i,\theta_j) &= k_q^2(r_1\cos\theta_i - r_2\cos\theta_j)^2\\
	&\qquad + k_p^2(r_1\sin\theta_i - r_2\sin\theta_j)^2,
\end{split}
\end{align}
where we label points on the isosurfaces by their angle $\theta$ on the original circular isosurfaces. The SHO optimal transport plan $\Pi$ takes $\theta$ on isosurface 1 to the point with the same angle $\theta$ on isosurface 2, and $\Pi$ for the SHO is invariant to coordinate rescaling (Appendix \ref{apx:proof}). Therefore, the total transport cost is
\begin{equation}
	C = \frac{1}{2\pi}\int_0^{2\pi}c(\theta,\theta)\,\mathrm{d}\theta = \frac{k_q^2+k_p^2}{2}(r_1 - r_2)^2,
\end{equation}
so the 2-Wasserstein distance is 
\begin{equation}
    \sqrt{C} = \sqrt{(k_q^2+k_p^2)/2}\,\lvert r_1 - r_2\rvert \propto \lvert r_1 - r_2\rvert,
\end{equation}
i.e.\ the same result modulo a constant factor. While this is not a general result, we find that our approach is often fairly robust to such changes, including the extreme case of scaling some phase space coordinates all the way down to zero resulting in a partially observed phase space (Appendix \ref{apx:robust}).

\subsection{Diffusion Maps: Extracting the Conserved Energy}

Once we have pairwise distances in the isosurface shape space, we can use diffusion maps to study the resulting manifold of isosurface shapes. With sufficient samples, the operator constructed by the diffusion map should converge to the Laplace--Beltrami operator on the manifold. For the SHO, the isosurface shape space is isomorphic to $\mathbb R^+$ with each circular isosurface mapped to its radius. If we sample trajectories with radii $r \in (0, \sqrt{2E_0})$ for some maximum energy $E_0$, then the manifold is a real line segment, and the resulting Laplacian operator (with open boundary conditions) has eigenvalues $\lambda_n = \pi^2n^2/2E_0$ and corresponding eigenvectors $v_n(r) = \cos(\sqrt{\lambda_n}\,r)$. Therefore, the first eigenvector or embedding component
\begin{equation}\label{eq:v1}
    v_1(E) = \cos(\pi\sqrt{E/E_0})   
\end{equation}
successfully encodes the conserved energy and is, in fact, a monotonic function of the energy.

\section{Numerical Experiments}

To demonstrate and empirically test our method for discovering conservation laws, we generate datasets from a wide range of dynamical systems, each consisting of randomly sampled trajectories with different initial conditions and the corresponding conserved quantities. Note that we use the dimensionless form of each dynamical system. All of the code necessary for reproducing our results is available at \url{https://github.com/peterparity/conservation-laws-manifold-learning}.

\subsection{Simple Harmonic Oscillator}\label{sec:sho}

\begin{figure}[t!]
    \centering
    \includegraphics[scale=0.88]{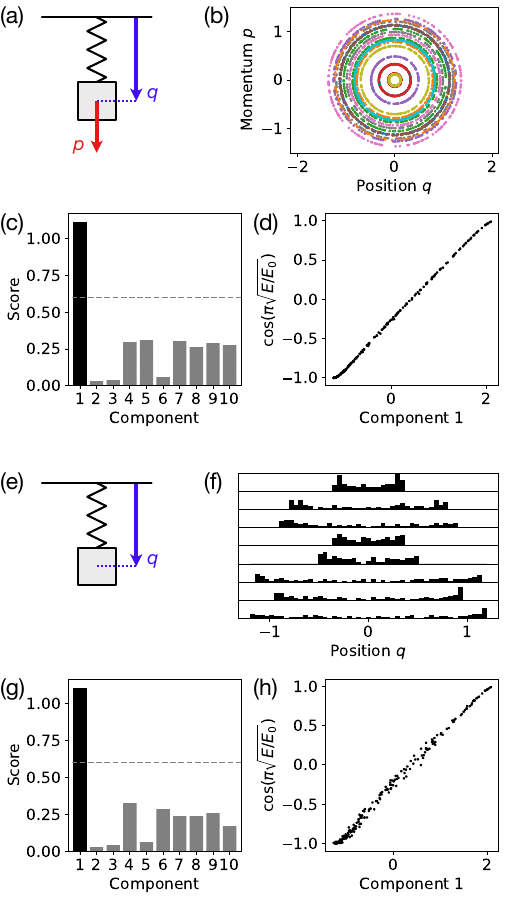}
    \caption{\textbf{Identifying the conserved energy for the simple harmonic oscillator (SHO).} (a) The SHO has two degrees of freedom: position $q$ and momentum $p$. (b) Sample trajectories from the SHO dataset show sample points plotted in the 2D phase space $(q, p)$. (c) The heuristic score (with cutoff 0.6) correctly identifies that the first embedding component extracted by the diffusion map is the only relevant component. (d) The extracted first component closely matches the analytically predicted first component (Eq.~\ref{eq:v1}) for the SHO ($R^2=0.9995$). (e) Next, consider the SHO dataset with a partially observed phase space containing position only. (f) For each sample trajectory, the sample points are shown as a histogram. (g) The heuristic score is still able to identify the first component as relevant, and (h) this first component matches the analytic prediction ($R^2=0.9961$).}
    \label{fig:sho}
\end{figure}

We first numerically test our analytic result for the SHO and obtain good agreement (Fig.~\ref{fig:sho}) using both the default scaling $k_q = k_p = 1$ (Figs.~\ref{fig:sho}a--d) as well as the position only scaling $k_q = 1$, $k_p = 0$ (Figs.~\ref{fig:sho}e--h), which effectively reduces the dimension of the phase space. A linear fit of the first embedding component from the diffusion map with the analytically predicted component (Eq.~\ref{eq:v1}) achieves a correlation coefficient of $R^2 = 0.9995$ for the default scaling and $R^2=0.9961$ for the position only scaling. We also verify that the heuristic score (Appendix \ref{apx:score}) accurately determines that there is only one relevant embedding component (Figs.~\ref{fig:sho}c,g), which corresponds to the conserved energy.

\subsection{Simple Pendulum}\label{sec:pendulum}

\begin{figure}[t!]
    \centering
    \includegraphics[scale=0.88]{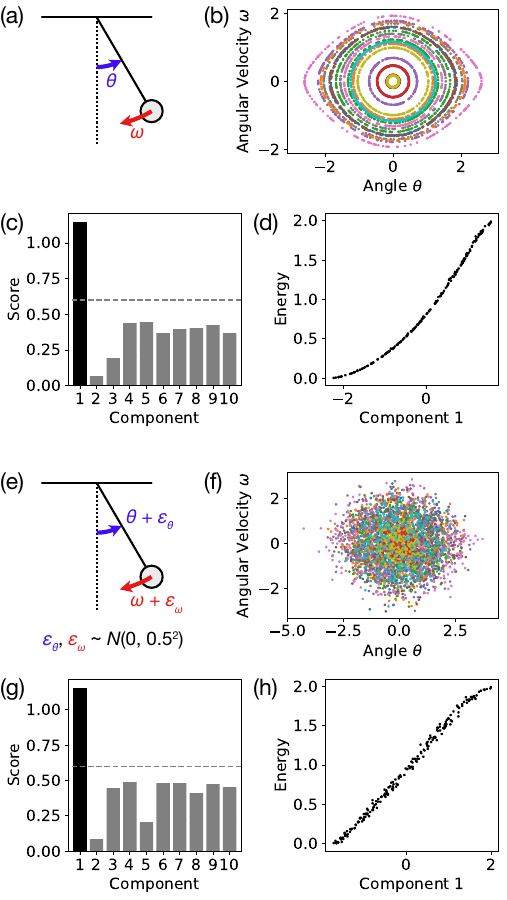}
    \caption{\textbf{Identifying the conserved energy for the simple pendulum.} (a) The simple pendulum has two degrees of freedom: angular position $\theta$ and angular velocity $\omega$. (b) Sample trajectories show sample points plotted in the 2D phase space $(\theta, \omega)$. (c) The heuristic score (with cutoff 0.6) correctly identifies that the first embedding component extracted by the diffusion map is the only relevant component, and (d) the extracted first component is monotonically related to the energy (rank correlation $\rho=0.9997$). (e, f) With the addition of $\sigma = 0.5$ Gaussian noise to simulate measurement noise, (g) the heuristic score is still able to identify the first component as relevant, and (h) this first component corresponds well to the energy ($\rho=0.9978$).}
    \label{fig:pendulum}
\end{figure}

To demonstrate our method on a simple nonlinear dynamical system, we analyze a simple pendulum that has a 2D phase space consisting of the angle $\theta$ and angular momentum $\omega$ (Fig.~\ref{fig:pendulum}a). The equations of motion are
\begin{align}
\begin{split}
    \frac{\mathrm{d}\omega}{\mathrm{d}t} &= -\sin\theta\\
    \frac{\mathrm{d}\theta}{\mathrm{d}t} &= \omega.
\end{split}
\end{align}
This system has a single scalar conserved quantity
\begin{equation}
    E = \frac{1}{2}\omega^2 + (1-\cos\theta)
\end{equation}
corresponding to the total energy of the pendulum, so the trajectories form 1D orbits in phase space (Fig.~\ref{fig:pendulum}b).

Our method is able to correctly determine that there is only a single conserved quantity (Fig.~\ref{fig:pendulum}c) corresponding to the energy of the pendulum (Fig.~\ref{fig:pendulum}d). The single extracted embedding component is monotonically related to the energy with Spearman's rank correlation coefficient $\rho = 0.9997$. We are also able to achieve similar results ($\rho = 0.9978$) with a high level of Gaussian noise (standard deviation $\sigma = 0.5$) added to the raw trajectory data (Figs.~\ref{fig:pendulum}e--h), showing that our approach is quite robust to measurement noise.

\subsection{Planar Gravitational Dynamics}\label{sec:orbits}

\begin{figure}[t]
    \centering
    \includegraphics[scale=0.88]{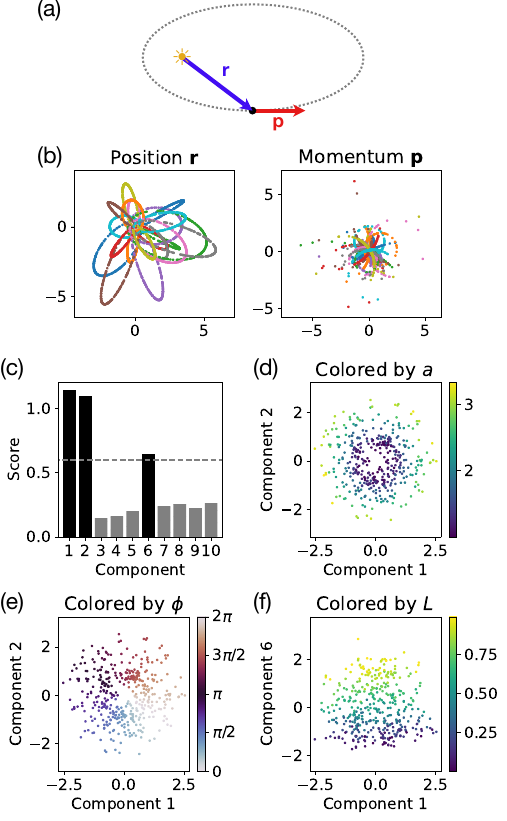}
    \caption{\textbf{Identifying conserved quantities for planar gravitational dynamics.} (a) Planar gravitational dynamics has four degrees of freedom: position vector $\mathbf{r}$ and momentum vector $\mathbf{p}$. (b) Sample trajectories show sample points plotted in 2D slices of the 4D phase space consisting of position $\mathbf{r}$ and momentum $\mathbf{p}$. (c) The heuristic score (with cutoff 0.6) identifies three relevant embedding components corresponding to the three independent conserved quantities. (d, e) Components 1 and 2 embed the semi-major axis vector $\mathbf{a}$ with magnitude $a = -1/2E$ related to the energy and orientation given by the angle $\phi$. (f) Component 6 corresponds to the angular momentum $L$.}
    \label{fig:orbits}
\end{figure}

To test our method on a system with multiple conserved quantities, we simulate the gravitational system of a planet orbiting a star with much greater mass (Fig.~\ref{fig:orbits}a). We fix the orbits to all lie in a 2D plane, giving us an effectively 4D phase space. The resulting equations of motion are
\begin{align}
\begin{split}
    \frac{\mathrm{d}\mathbf{r}}{\mathrm{d}t} &= \mathbf{p}\\
    \frac{\mathrm{d}\mathbf{p}}{\mathrm{d}t} &= -\frac{\hat{\mathbf{r}}}{\lvert\mathbf{r}\rvert^2}.
\end{split}
\end{align}
This system has one scalar and two vector conserved quantities
\begin{align}
\begin{split}
    E &= \frac{\mathbf{p}^2}{2} - \frac{1}{\lvert\mathbf{r}\rvert}\\
    \mathbf{L} &= \mathbf{r}\times\mathbf{p}\\
    \mathbf{A} &= \mathbf{p}\times\mathbf{L}-\hat{\mathbf{r}},
\end{split}
\end{align}
which, in our 4D phase space, reduces to three scalar conserved quantities: the total energy $E$ (or equivalently, the semi-major axis $a = -1/2E$), the angular momentum $L = \lvert\mathbf{L}\rvert$, and the orbital orientation angle $\phi$, which is the angle of the LRL vector $\mathbf{A}$ relative to the $x$-axis. As a result, the trajectories also form 1D orbits in the phase space (Fig.~\ref{fig:orbits}b).

Our approach accurately identifies the three conserved quantities (Fig.~\ref{fig:orbits}c), and the extracted embedding corresponds most directly to the geometric features of the orbits (Figs.~\ref{fig:orbits}d--f). The first two components embed the semi-major axis vector $\mathbf{a} = (a\cos\phi, a\sin\phi)$ with magnitude given by the semi-major axis $a=-1/2E$, which is related to the energy $E$, and orientation given by the orientation angle $\phi$ of the elliptical orbit (Figs.~\ref{fig:orbits}d,e). The third relevant component (component 6) embeds the angular momentum $L$ (Fig.~\ref{fig:orbits}f). See Appendix \ref{apx:scorecutoff} for details on choosing a cutoff to identify the relevant components. A linear fit of the identified relevant embedding components with $a\cos\phi$ ($a\sin\phi$) has $R^2= 0.987$ ($R^2=0.986$) and rank correlation $\rho = 0.994$ ($\rho=0.992$). A similar linear fit with $L$ has $R^2 = 0.927$ and $\rho=0.970$.

This example demonstrates that, for a system with multiple conserved quantities, the ground metric for optimal transport controls the relative scale of each conserved quantity in the extracted embedding. In this case, the geometry of the shape space $\mathcal C$ is dominated by changes in the semi-major axis $a$ and orientation angle $\phi$, whereas changes in the angular momentum $L$, which controls the eccentricity of the orbit, play a more minor role and thus appear in a later embedding component with a lower score (Fig.~\ref{fig:orbits}c).

\subsection{Double Pendulum}\label{sec:double}

\begin{figure}[t!]
    \centering
    \includegraphics[scale=0.88]{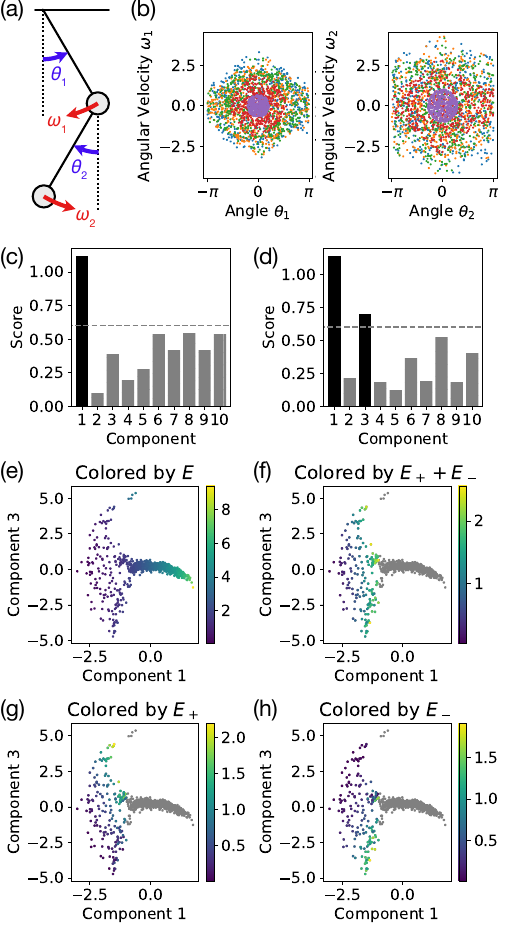}
    \caption{\textbf{Identifying conserved quantities for the double pendulum.} (a) The double pendulum has four degrees of freedom: angular positions $\theta_1,\theta_2$ and angular velocities $\omega_1,\omega_2$. (b) Sample trajectories show sample points plotted in 2D slices of the 4D phase space. (c) The heuristic score (with cutoff 0.6) identifies one relevant embedding component corresponding to (e) the total energy $E$. (d) However, if we restrict the embedding to trajectories with first component $v_1 < -1$ (i.e.\ low energy trajectories) and renormalize the embedding, we find (f--h) two conserved quantities corresponding to the energies $E_\pm$ of the two decoupled low energy modes. The gray points in Figures \ref{fig:double}f--h correspond to the high energy trajectories (first component $v_1 > -1$) which are not relevant when considering the low energy non-chaotic phase of the double pendulum.}
    \label{fig:double}
\end{figure}

To test our approach on a non-integrable system with higher dimensional isosurfaces, we study the classic double pendulum system (Fig.~\ref{fig:double}a) with unit masses and unit length pendulum arms. This system has a 4D phase space, consisting of the angles $\theta_1, \theta_2$ and the angular velocities $\omega_1,\omega_2$ of the two pendulums (Fig.~\ref{fig:double}b), and only has a single scalar conserved quantity
\begin{equation}
    E = \omega_1^2+\frac{1}{2}\omega_2^2 + \omega_1\omega_2\cos(\theta_1-\theta_2) - 2\cos\theta_1 - \cos\theta_2
\end{equation}
corresponding to the total energy. However, the double pendulum system has both chaotic and non-chaotic phases. In particular, at high energies, the system is chaotic and only conserves the total energy, while at low energies, the system behaves more like two coupled harmonic oscillators with two independent (approximately) conserved energies
\begin{align}
\begin{split}
    E_\pm &= \frac{1}{8} \left[4\theta_1^2  + 2\theta_2^2 \pm \sqrt{2}\,\theta_1\theta_2 \right.\\
    &\qquad + \left(2\pm\sqrt{2}\right)\left(2\omega_1^2+\omega_2^2\right)\\
    &\qquad \left. + 4
   \left(1\pm\sqrt{2}\right)\omega_1\omega_2\right]
\end{split}
\end{align}
corresponding to the two modes of the coupled oscillator system. Therefore, we expect to see two distinct phases in our extracted embedding: one with a single conserved quantity $E$ at high energy and another with two approximately conserved quantities $E_\pm$ at low energy, which approximately sum to $E \approx E_+ + E_-$.

At first glance, it appears as though our method has only identified a single relevant component corresponding to the conserved total energy $E$ (Figs.~\ref{fig:double}c,e) with rank correlation $\rho = 0.996$. However, if we restrict ourselves to low energy trajectories with first embedding component $v_1 < -1$, we find that there is a region of the shape space that is two-dimensional, corresponding to the two independently conserved energies $E_\pm$ of the low energy non-chaotic phase where the double pendulum behaves like a coupled oscillator system with two distinct modes. For the low energy trajectories, a linear fit of the now two relevant components with $E_+$ ($E_-$) has rank correlation $\rho = 0.919$ ($\rho=0.937$). If we restrict ourselves to even lower energy trajectories with $v_1 < -2$, a similar linear fit for $E_+$ ($E_-$) has rank correlation $\rho = 0.990$ ($\rho = 0.989$).

This analysis of the double pendulum shows that our method can still provide significant insight into complex dynamical systems with multiple phases involving varying numbers of conserved quantities. This manifests itself as manifolds of different dimensions in shape space that are stitched together at phase transitions, presenting a significant challenge for most manifold learning methods. In this example, this difficulty is reflected in the performance of the heuristic score (Fig.~\ref{fig:double}c,d), which has trouble telling whether the embedding is one or two dimensional precisely because it is a combination of both a one and two dimensional manifold. The embedding, on the other hand, remains very informative despite the sudden change in dimensionality and allows us to identify interesting features of the system, such as nonlinear periodic orbits (see Appendix \ref{apx:doublependulumorbit}). The effectiveness of diffusion maps when handling these complex situations has been previously observed in parameter reduction applications \cite{HOLIDAY2019419} and is worth studying in more detail in the future.

\subsection{Oscillating Turing Patterns}\label{sec:turing}

Next, we consider an oscillating Turing pattern system that is both dissipative and has a much higher dimensional phase space than our previous examples. In particular, we study the Barrio--Varea--Arag\'{o}n--Maini (BVAM) model \cite{BARRIO1999483,PhysRevE.86.026201}
\begin{align}
\begin{split}
    \frac{\partial u}{\partial t} &= D\frac{\partial^2 u}{\partial x^2} + u - v - Cuv - uv^2\\
    \frac{\partial v}{\partial t} &=\frac{\partial^2 v}{\partial x^2} -\frac{3}{2}v + Hu + Cuv + uv^2
\end{split}
\end{align}
with $D=0.08$, $C=-1.5$, and $H=3$, following Arag\'{o}n
\textit{et al.} \cite{PhysRevE.86.026201} who showed that this set of parameters results in a spatial Turing pattern that also exhibits chaotic oscillating temporal dynamics, on a periodic domain with size $8$. The phase space of the BVAM system consists of two functions $u(x)$ and $v(x)$\footnote{In our method, each trajectory $[u(x,t_i), v(x,t_i)], i \in \{1,2,\ldots, N\}$ is treated as an unordered set of sample points in phase space, so we refer to the phase space as $[u(x), v(x)]$ in a slight abuse of notation.} which we discretize on a mesh of size $50$, giving us an effective phase space dimension of $100$. Because this system is dissipative, we will focus on characterizing the long term behavior of the dynamics, i.e.\ the oscillating Turing pattern, which appears to have a conserved spatial phase $\eta$ for our chosen set of parameters corresponding to the spatial position of the Turing pattern. In the language of dynamical systems, $\eta$ parameterizes a continuous set of attractors for this dissipative system.

\begin{figure}[t]
    \centering
    \includegraphics[scale=0.88]{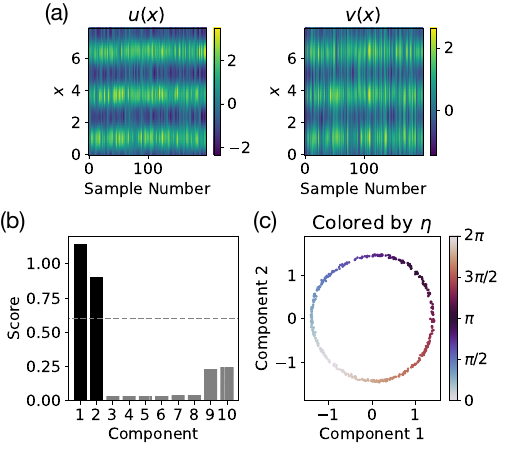}
    \caption{\textbf{Identifying the conserved spatial phase for the oscillating Turing pattern system.} (a) An example trajectory, with randomly sampled states $u(x)$ and $v(x)$ plotted, illustrates the high dimensional nature of the problem. (b) The heuristic score (with cutoff 0.6) identifies two relevant components, but on further examination, (c) we see that there is just a single conserved angle, corresponding to the spatial phase $\eta$ of the Turing pattern, that needs to be embedded in two dimensions due to its topology.}
    \label{fig:turing}
\end{figure}

Our method successfully identifies the spatial phase $\eta$ but embeds the angle as a circle in a 2D embedding space (Fig.~\ref{fig:turing})---a result of the periodic topology of $\eta$. While this shows that the number of relevant components determined by our heuristic score may not always match the true manifold dimensionality, such cases are often easily identified by examining the components directly (Fig.~\ref{fig:turing}c) or by cross checking with an intrinsic dimensionality estimator \cite{https://doi.org/10.48550/arxiv.2106.04018}. A linear fit of the two relevant components with $\cos\eta$ ($\sin\eta$) has $R^2 = 0.9991$ ($R^2 = 0.9997$) and $\rho = 0.9993$ ($\rho = 0.9992$). This example both tests our method on a high dimensional phase space and demonstrates how our approach can be applied to dissipative systems to study long term behavior.

\subsection{Korteweg--De Vries Equation}\label{sec:kdv}

\begin{figure}[t]
    \centering
    \includegraphics[scale=0.88]{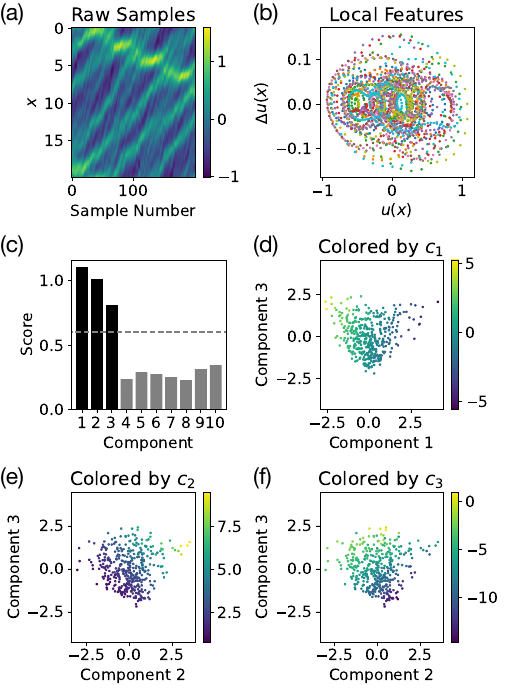}
    \caption{\textbf{Identifying three local conserved quantities of the Korteweg--De Vries (KdV) equation.} (a) An example trajectory from the KdV dataset shows the high dimensional raw sampled states $u(x)$. (b) To focus on local conserved quantities, we extract a distribution of the local features $u(x),\Delta u(x)$ from the raw states, removing the explicit spatial label. The plot shows the local feature distributions for a few sample states. (c) The heuristic score (with cutoff 0.6) correctly identifies three relevant components corresponding to (d--f) the three local conserved quantities (Eq.~\ref{eq:kdv}).}
    \label{fig:kdv}
\end{figure}

For many spatiotemporal dynamical systems, the conservation laws are local in nature. Locality can significantly simplify the analysis of the conserved quantities and suggests a way to restrict the type of conserved quantities identified by our method. Specifically, we can adapt our approach to focus on local conserved quantities by replacing the raw states (Fig.~\ref{fig:kdv}a) by a distribution of local features (Fig.~\ref{fig:kdv}b), removing the explicit spatial label and providing a fully translation invariant representation of the state. Then, instead of using the Euclidean metric in the original phase space, we use the energy distance \cite{https://doi.org/10.1002/wics.1375,pmlr-v89-feydy19a} between the distributions of local features as the ground metric for optimal transport.

To demonstrate this method for identifying local conserved quantities, we consider the Korteweg--De Vries (KdV) equation
\begin{equation}
    \frac{\partial u}{\partial t} = -\frac{\partial^3 u}{\partial x^3} - 6u\frac{\partial u}{\partial x}.
\end{equation}
The KdV equation is fully integrable \cite{PhysRevLett.19.1095} and has infinitely many conserved quantities \cite{doi:10.1063/1.1664701}, the most robust of which are the most local conserved quantities expressible in terms of low order spatial derivatives. To focus on these robust local conserved quantities, we use finite differences (i.e.\ $u(x), \Delta u(x) = u(x+\Delta x) - u(x), \Delta^2 u(x), \ldots$) as our local features, allowing us to restrict the spatial derivative order of the identified conserved quantities. In this experiment, we only take $u(x)$ and $\Delta u(x)$, meaning that the identified local conserved quantities will only contain up to first order spatial derivatives. For the KdV equation, there are three such local conserved quantities:
\begin{align}
\begin{split}
    c_1 &= \int_0^l u\,\mathrm{d}x\\
    c_2 &= \int_0^l u^2\,\mathrm{d}x\\
    c_3 &= \int_0^l \left[u^3 - \frac{1}{2}\left(\frac{\partial u}{\partial x}\right)^2\right]\,\mathrm{d}x,
\end{split}\label{eq:kdv}
\end{align}
where $c_1$ and $c_2$ are often identified as ``momentum'' and ``energy'', respectively \cite{doi:10.1063/1.1664701}. These local conserved quantities also have direct analogues in generalized KdV-type equations, hinting at their robustness \cite{1937-1632_2018_4_607}.

Our method successfully identifies three relevant components (Fig.~\ref{fig:kdv}c) corresponding to (d--f) the three local conserved quantities (Eq.~\ref{eq:kdv}). Linear fits of these components to $c_1$, $c_2$, and $c_3$ have rank correlations $\rho = 0.995$, $0.994$, and $0.985$, respectively. This result shows how our approach can be adapted to incorporate known structure, such as locality and translation symmetry, in applications to complex high dimensional dynamical systems.

\section{Discussion}

We have proposed a non-parametric manifold learning method for discovering conservation laws, tested our method on a wide variety of dynamical systems---including complex chaotic systems with multiple phases and high dimensional spatiotemporal dynamics---and also shown how to adapt our approach to incorporate additional structure such as locality and translation symmetry. While our experiments use dynamical systems with known conserved quantities in order to validate our approach, our method does not require any \emph{a priori} information about the conserved quantities. Our method also does not assume or construct an explicit model for the system nor require accurate time information like previous approaches \cite{8618963,https://doi.org/10.48550/arxiv.2203.12610}, only relying on the ergodicity of the dynamics modulo the conservation laws (Sec.~\ref{sec:ergodicity}). As a result, our approach is also quite robust to measurement noise and can often deal with missing information such as a partially observed phase space (Figs.~\ref{fig:sho}e--h, Figs.~\ref{fig:pendulum}e--h, Appendix \ref{apx:robust}).

Compared with recently proposed deep learning-based methods \cite{PhysRevResearch.2.033499,PhysRevResearch.3.L042035}, our approach is substantially more interpretable since it relies on explicit geometric constructions and well-studied manifold learning methods that directly determine the geometry of the shape space $\mathcal C$ and, therefore, the identified conserved quantities. This is reflected in our ability to explicitly derive the expected result for the simple harmonic oscillator (Sec.~\ref{sec:analytic}), as well as in the identified conserved quantities in many of our experiments. For example, the embedding of the semi-major axis vector in the planar gravitational dynamics experiment (Sec.~\ref{sec:orbits}) stems directly from the elliptical geometry of the orbits and their orientation in phase space, which is captured by the Euclidean ground metric and lifted into shape space by optimal transport. Our method also correctly captures the subtleties of the double pendulum system (Sec.~\ref{sec:double}) by providing an embedding that shows both a 1D manifold at high energies and a 2D manifold at low energies---a difficult prospect for deep learning approaches that try to explicitly fit conserved quantities. In addition, we empirically find that our method outperforms existing direct fitting approaches \cite{PhysRevResearch.2.033499,PhysRevResearch.3.L042035}. See Appendix \ref{apx:comparison} for a comparison benchmark using our planar gravitational dynamics dataset.

Our manifold learning approach to identifying conserved quantities provides a new way to analyze data from complex dynamical systems and uncover useful conservation laws that will ultimately improve our understanding of these systems as well as aid in developing predictive models that accurately capture long term behavior. While our method does not provide explicit symbolic expressions for the conserved quantities (which may not exist in many cases), we do obtain a full set of independent conserved quantities, meaning that any other conserved quantity will be a function of the discovered ones. Our method also serves as a strong non-parametric baseline for future methods that aim to discover conservation laws from data. Finally, we believe that similar combinations of optimal transport and manifold learning have the potential to be applied to a wide variety of other problems that also rely on geometrically characterizing families of distributions, and we hope to investigate such applications in the near future.

\backmatter





\bmhead{Acknowledgments}

We would like to acknowledge useful discussions with Justin Solomon, Ziming Liu, Andrew Ma, Samuel Kim, Charlotte Loh, and Ruba Houssami. P.Y.L. gratefully acknowledges the support of the Eric and Wendy Schmidt AI in Science Postdoctoral Fellowship, a Schmidt Futures program. This research is supported in part by the U.S. Department of Defense through the National Defense Science \& Engineering Graduate Fellowship Program; the National Science Foundation under Cooperative Agreement PHY-2019786 (The NSF AI Institute for Artificial Intelligence and Fundamental Interactions, \url{http://iaifi.org/}); the U.S. Army Research Office through the Institute for Soldier Nanotechnologies at MIT under Collaborative Agreement Number W911NF-18-2-0048; the Air Force Office of Scientific Research under the award number FA9550-21-1-0317; and the United States Air Force Research Laboratory and the United States Air Force Artificial Intelligence Accelerator under Cooperative Agreement Number FA8750-19-2-1000. The views and conclusions contained in this document are those of the authors and should not be interpreted as representing the official policies, either expressed or implied, of the United States Air Force or the U.S.\ Government. The U.S.\ Government is authorized to reproduce and distribute reprints for Government purposes notwithstanding any copyright notation herein.

\section*{Declarations}

\bmhead{Competing interests}
The authors declare no competing interests.

\bmhead{Data availability}
All the datasets used in this study can be generated using the publicly available data generation scripts provided at \url{https://github.com/peterparity/conservation-laws-manifold-learning}.

\bmhead{Code availability}
All the code necessary for reproducing our results is publicly available at \url{https://github.com/peterparity/conservation-laws-manifold-learning}.

\bmhead{Author contributions}
All three authors contributed to the conception, design, and development of the proposed method. P.Y.L. implemented, tested, and refined the method and also generated the datasets. The manuscript was written by P.Y.L. with support from R.D. and M.S.

\begin{appendices}

\section{Heuristic Score for a Minimal Diffusion Maps Embedding}\label{apx:score}

Traditionally, diffusion maps \cite{Coifman2005} and Laplacian eigenmaps \cite{6789755} leave the embedding dimension $n$ as a hyperparameter and simply use the eigenvectors corresponding to the $n$ smallest eigenvalues to construct the embedding. In practice, the embedding dimension $n$ is often chosen for convenience (e.g.\ in visualization applications) or by examining the eigenvalues $\lambda_i$ and looking for a sharp increase in the magnitude of the eigenvalues that would separate the signal from the noise. Because identifying the number of conservation laws is an important step in our approach, we refine this heuristic by directly computing an approximate length scale
\begin{equation}\label{eq:length}
    l_i  = \sqrt{-\epsilon/\log(1-\lambda_i)},
\end{equation}
where $\epsilon$ is the scale factor from the Gaussian kernel (Eq.~\ref{eq:kernel}) used to construct the Laplacian matrix $L$. We derive this length scale by considering the normalized kernel $I-L$ to be an approximation of the heat kernel $\exp(\epsilon\Delta)$, implying that the length scales $l_i$ associated with the Laplace--Beltrami operator $\Delta$ are given by
\begin{equation}
    \exp(\epsilon\Delta) = I-L \implies \exp(-\epsilon/l_i^2) = 1 - \lambda_i.
\end{equation}
We then divide by $l_1$ to obtain the relative length scale
\begin{equation}
    \frac{l_i}{l_1} = \sqrt{\frac{\log(1-\lambda_1)}{\log(1-\lambda_i)}},
\end{equation}
which can be used as a heuristic measure of relevance---components with a small relative length scale are more likely to be noise. Compared with directly using the eigenvalues $\lambda_i$, we find this heuristic to be less sensitive to the choice of $\epsilon$ in the kernel.

\begin{figure}[t]
    \centering
    \includegraphics[scale=0.88]{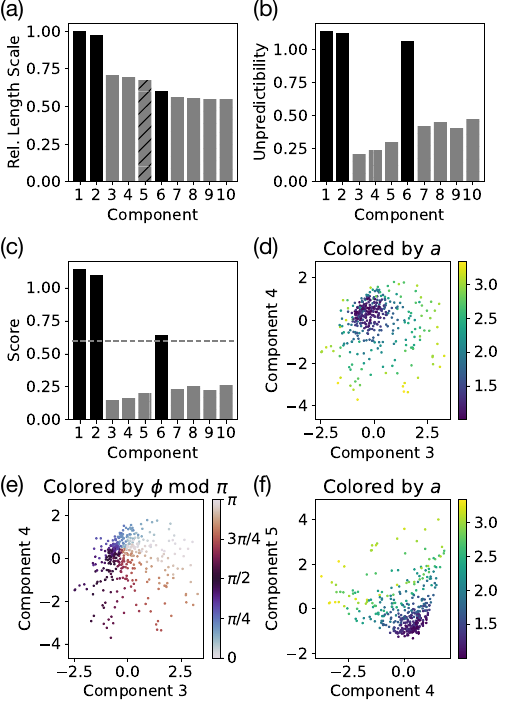}
    \caption{\textbf{Breakdown of the heuristic score and illustration of redundant embedding components from the planar gravitational dynamics experiment.} (a) The relative length scale $l_i/l_1$ for each embedding component is computed from the corresponding eigenvalue $\lambda_i$ of the Laplacian matrix (Eq.~\ref{eq:length}). (b) The unpredictability measure $m_i$ for each component is computed using a nearest neighbor estimator \cite{pfau2018minimally}. (c) The combined score $m_il_i/l_1$ is the product of the relative length scale and the unpredictability measure. (d--f) The components 3, 4, and 5 are identified by the unpredictability measure as redundant. If we examine these three components, we find that they together embed a second order angular mode of components 1 and 2 (Figs.~\ref{fig:orbits}d,e). In particular, the embedding is shaped like the surface of a cone with the height (or radial distance) roughly corresponding to the semi-major axis $a$ and the angle around the cone corresponding to $\phi\ \mathrm{mod}\ \pi$, a second order mode of the orientation angle $\phi$.}
    \label{fig:score}
\end{figure}

In addition to noise, there is the common problem of redundant embedding components that stem from the structure of the Laplacian operator: higher order modes of previous eigenvectors often appear before more informative eigenvectors corresponding to new manifold directions. This problem is clearly illustrated in the planar gravitational dynamics experiment (Sec.~\ref{sec:orbits}), where components 3, 4, and 5 are all redundant with components 1 and 2 but component 6 is a new and relevant conserved quantity (Figs.~\ref{fig:score}d--f). To address this issue, the key observation is that, while all components of the diffusion map are linearly independent, redundant components are still predictable (via a nonlinear function) from previous components. Therefore, we require a measure of ``unpredictability'' that allows us to identify redundancies. We choose the heuristic $m_i$ proposed by Pfau
and Burgess \cite{pfau2018minimally} that uses a nearest neighbor estimator (using 5 nearest neighbors) to determine whether a new embedding component is too predictable and therefore redundant. Alternative methods for dealing with these redundant components, also called repeated eigendirections or higher harmonics, have been proposed that use local linear regression to detect redundant components \cite{DSILVA2018759} or adapt the diffusion kernel to be anisotropic for chosen components \cite{vonLindheim2018}.

Our final heuristic score (Fig.~\ref{fig:score}c)
\begin{equation}\label{eqn:score}
    s_i = m_il_i/l_1
\end{equation}
is the product of the relative length scale $l_i/l_1$ (Fig.~\ref{fig:score}a) and the unpredictability measure $m_i$ (Fig.~\ref{fig:score}b).
We find this simple combined score performs well for identifying relevant embedding components by removing both noise components as well as redundant components.

\subsection{Choosing a Score Cutoff}\label{apx:scorecutoff}

\begin{figure}[t]
    \centering
    \includegraphics[scale=0.88]{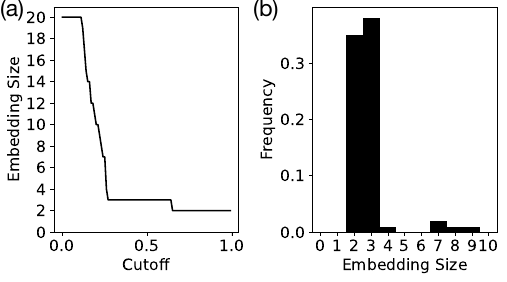}
    \caption{\textbf{Example from the planar gravitational dynamics experiment of identifying the number of conserved quantities (i.e.\ the embedding size).} (a) Sweeping the cutoff value from 0 to 1, we find plateaus indicating robustness at embedding size 2 and 3. Note that there is a spurious plateau at the maximum embedding size 20. (b) A histogram of the embedding sizes confirms that the number of conserved quantities is likely to be 3.}
    \label{fig:cutoff}
\end{figure}

To use the heuristic score to identify the number of conserved quantities and construct a minimal embedding, we require a score cutoff to separate relevant components that we keep in our embedding from irrelevant components that we discard. To choose this cutoff, we sweep cutoff values in the interval $[0, 1]$, compute the embedding size (i.e.\ the number of relevant components) based on the chosen cutoff, and then examine the result to identify a robust value for the cutoff (Fig.~\ref{fig:cutoff}). Specifically, we look for wide plateaus in the embedding size that indicate robustness to the value of the cutoff and find that a cutoff of 0.6 works well in all of our experiments. In practice, we would treat a cutoff of 0.6 as a good starting point but recommend analyzing a range of cutoff values to find a robust choice of cutoff, as illustrated in Fig.~\ref{fig:cutoff}.

\section{Additional Method Details}\label{apx:details}

All of the code necessary for generating our datasets, applying our method, and reproducing our results is available at \url{https://github.com/peterparity/conservation-laws-manifold-learning}.

\subsection{Sinkhorn Algorithm}\label{apx:sinkhorn}

\subsubsection{Entropy Regularized Optimal Transport}

We compute an approximate 2-Wasserstein distance using the Sinkhorn algorithm \cite{NIPS2013_af21d0c9}, which solves an entropy regularized relaxation of the optimal transport problem
\begin{equation}\label{eq:discreteW2entropy}
    \widetilde{W}_2(\{\mathbf{x}_i\},\{\mathbf{y}_j\}) = \bigg(\min_{T} \sum_{i,j}T_{ij}C_{ij} - \gamma\,h(T)\bigg)^{1/2},
\end{equation}
where the cost matrix $C_{ij} = \lVert\mathbf{x}_i-\mathbf{y}_j\rVert^2$ and the entropy $h(T) = -\sum_{i,j}T_{ij}\log T_{ij}$. $\widetilde{W}_2$ reduces to the exact 2-Wasserstein distance $W_2$ (Eq.~\ref{eq:discreteW2}) as $\gamma \to 0$. For $\gamma > 0$, the entropy regularization introduces a smoothing bias that manifests as a nonzero ``self-distance'' $\widetilde{W}_2(\{\mathbf{x}_i\},\{\mathbf{x}_j\}) > 0$. This can be corrected by instead using the Sinkhorn divergence \cite{pmlr-v89-feydy19a,pmlr-v119-janati20a}
\begin{equation}
\begin{aligned}
    \overline{W}_2 =& \bigg(\widetilde{W}_2(\{\mathbf{x}_i\},\{\mathbf{y}_j\})^2\\
    &- \frac{\widetilde{W}_2(\{\mathbf{x}_i\},\{\mathbf{x}_j\})^2 + \widetilde{W}_2(\{\mathbf{y}_i\},\{\mathbf{y}_j\})^2}{2}\bigg)^{1/2}
\end{aligned}
\end{equation}
as our estimate for the 2-Wasserstein distance, which explicitly subtracts off this self-distance. In our experiments, we use a convergence threshold of $\varepsilon = 0.01$ and a decaying entropy regularization parameter $\gamma$ that starts at $10.0$ and decays by a factor of $0.995$ at each step until it reaches a target of $0.1$.

Note that the Sinkhorn algorithm is a general approach for solving entropy-regularized optimal transport and is equally good at approximating a 1-Wasserstein distance (or Earth mover's distance). We experimented with using 1-Wasserstein vs.\ 2-Wasserstein distances and did not find a significant difference in the performance of our method.

\subsubsection{Time Complexity}

With $S$ samples per trajectory, the Sinkhorn algorithm solves the entropy regularized optimal transport problem in $\mathcal O(S^2\log S/\varepsilon^2)$ time for an $\varepsilon$-accurate solution \cite{NIPS2017_491442df,pmlr-v80-dvurechensky18a} using $\mathcal O(S)$ space (without explicit storing the cost matrix $C$ \cite{https://doi.org/10.48550/arxiv.2201.12324}). Therefore, the time complexity of computing approximate Wasserstein distances for all pairs of trajectories is $\mathcal O(N^2S^2\log S/\varepsilon^2)$ for a dataset containing $N$ total trajectories. This computation is currently the performance bottleneck of our approach (Appendix \ref{apx:complexity}) but is easily parallelized over multiple GPUs using the OTT-JAX library \cite{https://doi.org/10.48550/arxiv.2201.12324}.

\subsection{Diffusion Maps}

\subsubsection{Choice of Manifold Learning Method}

\begin{figure}[t]
    \centering
    \includegraphics[scale=0.88]{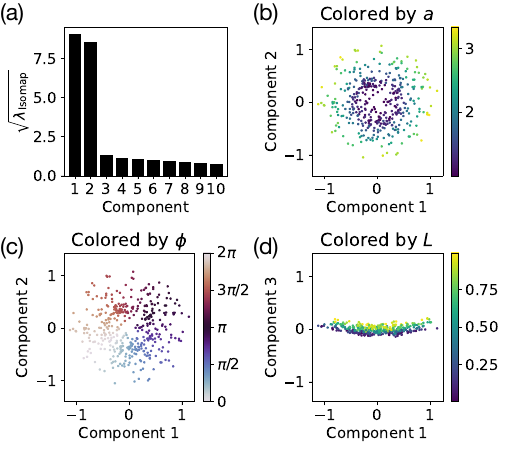}
    \caption{\textbf{Identifying conserved quantities for planar gravitational dynamics using Isomap.} (a) When applying Isomap, the effective length scale $\sqrt{\lambda_\mathrm{Isomap}}$ can be used to estimate manifold dimensionality but it only clearly identifies two out of the three conserved quantities. (b, c) Similarly to the diffusion map embedding, Isomap components 1 and 2 embed the semi-major axis vector $\mathbf{a}$ with magnitude $a = -1/2E$ related to the energy and orientation given by the angle $\phi$. These components have high rank correlation $\rho = 0.994$ ($\rho = 0.992$) with $a\cos\phi$ ($a\sin\phi$).  (d) Component 3 roughly corresponds to the angular momentum $L$ but is noisier than the embedding identified by diffusion maps (Fig.~\ref{fig:orbits}f) and has a much lower rank correlation $\rho = 0.723$ with $L$.}
    \label{fig:orbitsisomap}
\end{figure}

Unlike many standard manifold learning applications, our input to the manifold learning method is not a set of points in Euclidean space but rather a pairwise Wasserstein distance matrix. Many popular methods, such as local linear embedding \cite{doi:10.1126/science.290.5500.2323} and local tangent space alignment \cite{doi:10.1137/S1064827502419154}, explicitly require the data to be embedded in a Euclidean space and so are not applicable for our problem. In addition to diffusion maps, we also experimented with Isomap \cite{doi:10.1126/science.290.5500.2319}, which can also take a distance matrix as input. However, we found Isomap embeddings to be noisier and less reliable than diffusion map embeddings (e.g.~see Fig.~\ref{fig:orbitsisomap}). Diffusion maps \cite{Coifman2005} or Laplacian eigenmaps \cite{6789755} also provide an effective way to estimate manifold dimensionality and choose relevant embedding components (Appendix \ref{apx:score}).

\subsubsection{Gaussian Kernel Width}\label{apx:kernelwidth}

The primary hyperparameter in our diffusion maps algorithm is the width parameter $\epsilon$ of the Gaussian kernel (Eq.~\ref{eq:kernel}). Generally speaking, we should choose $\epsilon$ to be large enough to avoid noise induced by sparse sampling but small enough for the true heat kernel to be well approximated by the Gaussian kernel \cite{6789755}. We choose $\epsilon$ such that the corresponding standard deviation $\sigma = \sqrt{\epsilon/2}$ of the Gaussian kernel is equal to the maximum distance to the $k$th nearest neighbor.

In practice, especially when the relevant embedding components are well separated from the noise in terms of length scale (Eq.~\ref{eq:length}), we find the diffusion map to be fairly insensitive to the choice of $k$, which we generally set to $k=20$ nearest neighbors. The only exception is for the planar gravitational dynamics dataset, where we use $k=200$. This is because the angular momentum---the least prominent of the three conserved quantities---has a slightly poorer reconstruction for $k=20$ ($\rho = 0.910$) and gets pushed back to component 10 of the embedding. While it is still clearly identifiable from noise, it requires a lower heuristic score cutoff of around $\sim 0.4$ and is easier to miss.

\subsubsection{Noise Robustness}

To improve the noise robustness of our diffusion map, we follow Karoui and Wu \cite{10.1214/14-AOS1275} and replace the diagonal of the affinity matrix $M$ (Eq.~\ref{eq:affinity}) with zeros, i.e.
\begin{equation}
    M^*_{ij} = M_{ij} - M_{ii}I_{ij},
\end{equation}
before constructing the Laplacian matrix $L$. Because this induces an overall shift in the eigenvalues of the Laplacian that interacts poorly with our length scale heuristic (Eq.~\ref{eq:length}), we correct for this by subtracting off the normalized mean shift
\begin{equation}
    s = \frac{1}{N}\sum_{i=1}^N \left(M_{ii} / \sum_{j}M_{ij}\right)
\end{equation}
from the Laplacian matrix $L$ to obtain the corrected Laplacian
\begin{equation}
    L^*_{ij} = L_{ij} - sI_{ij},
\end{equation}
which we use to generate our embeddings.

\subsubsection{Time Complexity}

For a fixed number of embedding dimensions and a dense kernel matrix $K$ with $N\times N$ entries ($N$ being the number of trajectories), diffusion maps have time complexity $\mathcal O(N^2)$ and space complexity $\mathcal O(N^2)$. In our experiments, computing the diffusion map is very fast with run times under one second for every dataset we tested.

\subsubsection{Out-of-Sample Embedding}

One complication of manifold learning methods like diffusion maps is that they do not provide an explicit way to embed new out-of-sample data. A naive approach would be to rerun the diffusion map algorithm on a combined dataset consisting of the original data used to create the embedding and the new data. However, this does not retain the original embedding and, in some cases, may be computationally prohibitive. A popular approach that does retain the original embedding is the Nystr\"om method \cite{NIPS2003_cf059682}, which embeds a new point using its pairwise distances with the original data and scales as $\mathcal O(N)$. For even faster embedding, landmark diffusion maps offer a significant speed-up by choosing a small subset of $M\ll N$ landmark points to use during Nystr\"om out-of-sample embedding \cite{LONG2019190}. This also has the added benefit of a reduced memory footprint, since only the $M$ landmark points need to be retained for embedding.

\section{Langevin Harmonic Oscillator}\label{apx:langevinsho}

\begin{figure}[t!]
    \centering
    \includegraphics[scale=0.88]{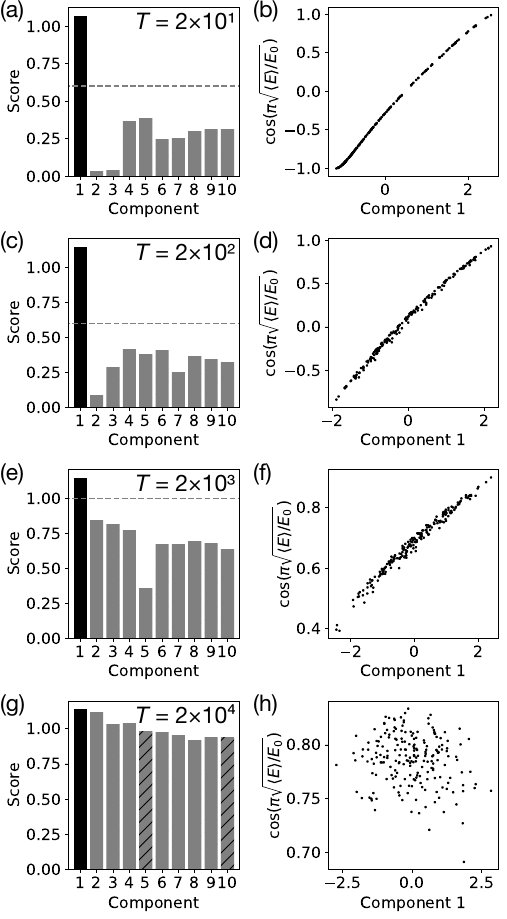}
    \caption{\textbf{Identifying conserved quantities for the Langevin harmonic oscillator over varying sampling times $T$.} (a) Over a short time scale $T=2\times 10^1$, the time-averaged energy $\langle E\rangle$ still clearly distinguishes the different trajectories, so our approach identifies a single (approximately) conserved quantity that corresponds to $\langle E\rangle$. (b) Component 1 of the embedding is highly correlated with $\langle E\rangle$ and still matches well with the theoretical result for the simple harmonic oscillator (Eq.~\ref{eq:v1}). (c,d) Over a slightly longer time scale $T=2\times 10^2$, we see a very similar result with a noisier fit between component 1 and $\langle E\rangle$. (e,f) For $T=2\times 10^3$, we start to see more ambiguity, with the identified component becoming significantly less prominent and an even noisier fit. (g,h) Over a long time scale $T=2\times 10^4$, the system has settled into a stationary distribution and no longer has any even approximately conserved quantities.}
    \label{fig:langevinsho}
\end{figure}
\begin{figure}[t!]
    \centering
    \includegraphics[scale=0.88]{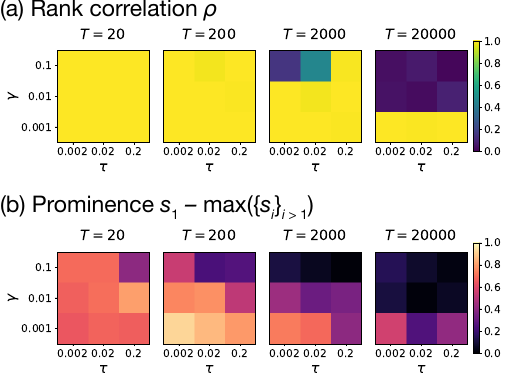}
    \caption{\textbf{Effect of varying $\gamma$ and $\tau$ on the identification of an approximate conservation law for the Langevin harmonic oscillator.} Varying $\gamma$, $\tau$, and the sampling time $T$, we show (a) the rank correlation $\rho$ of the first embedding component with the time-averaged energy $\langle E\rangle$ and (b) the prominence $s_1 - \max(\{s_i\}_{i>1})$ of the score of the first component $s_1$ over the highest score of the remaining components $s_i$.}
    \label{fig:langevinshovarying}
\end{figure}

To demonstrate our method on a simple example of a dynamical system with approximately conserved quantities at short time scales but no true conserved quantities, we consider an under-damped harmonic oscillator that is weakly coupled to a heat bath. The resulting dynamics are governed by the Langevin equations
\begin{equation}
\begin{aligned}
    \frac{\mathrm{d}q}{\mathrm{d}t} &= p\\
    \frac{\mathrm{d}p}{\mathrm{d}t} &= -q -\gamma p + \xi(t),
\end{aligned}
\end{equation}
where the damping $\gamma = 10^{-2}$. $\xi(t)$ is a Gaussian random process (i.e.\ Brownian motion) with zero mean and $\langle\xi(t)\xi(t')\rangle = 2\gamma \tau\,\delta(t-t')$. We set the temperature of the heat bath to be $\tau=2\times 10^{-2}$.
At short times $t \ll \gamma^{-1}$ and $t \ll (\gamma\tau)^{-1/2}$, the energy $E = (q^2+p^2)/2$ is approximately conserved. At long times $t \gg \gamma^{-1}$, all trajectories will sample a stationary distribution with variance $\langle q^2\rangle_0 = \tau$ and no conserved quantities.

We generate four datasets for this system, each with 200 trajectories and 200 samples over different periods of time, i.e.\ $t\in [0, T]$ with sampling time $T\in\{2\times 10^1,2\times 10^2,2\times 10^3,2\times 10^4\}$. This allows us to study how changing the sampling time $T$---the time scale over which we identify approximately conserved quantities---changes the embedding produced by our approach. Over short time scales, the time-averaged energy $\langle E\rangle$ is still a distinguishing feature of the trajectories and acts as an approximately conserved quantity (Figs.~\ref{fig:langevinsho}a--d). Over longer time scales, the energy becomes less and less relevant until, finally, the system reaches a stationary distribution (Figs.~\ref{fig:langevinsho}e--h).

\subsection{Separation of Time Scales and Dependence on $\gamma$ and $\tau$}

As previously noted, the energy of the Langevin harmonic oscillator is only conserved over time scales $T$ much less than the time scales associated with dissipation $\gamma^{-1}$ and random forcing $(\gamma\tau)^{-1/2}$. However, this approximate conservation law is only meaningful if the energy is roughly conserved over a time scale $T$ much greater than the period of oscillation, which is of order one in our units. In other words, approximate conservation laws only appear when there is a separation of time scales between fast conservative dynamics (e.g.~the oscillations of our harmonic oscillator) and slow non-conservative dynamics (e.g.~dissipation and forcing).

Thus, for $\gamma \ll 1$ and $\tau \ll 1$, we expect to have an approximately conserved energy at intermediate time scales $1 \ll T\ll \gamma^{-1}$ and $T \ll (\gamma\tau)^{-1/2}$. As $\gamma,\tau \to 1$, there is no longer an intermediate time scale for which the approximate conservation law holds. This is precisely what we see when we apply our method for identifying conservation laws to Langevin harmonic oscillators with varying $\gamma \in\{10^{-3}, 10^{-2},10^{-1}\}$ and $\tau\in\{2\times 10^{-3},2\times 10^{-2},2\times 10^{-1}\}$ using sampling times $T\in\{2\times 10^1,2\times 10^2,2\times 10^3,2\times 10^4\}$ (Fig.~\ref{fig:langevinshovarying}).

The rank correlation $\rho$ (Fig.~\ref{fig:langevinshovarying}a) shows the alignment of the first embedding component with the time-averaged energy $\langle E\rangle$, while the prominence $s_1 - \max(\{s_i\}_{i>1})$ (Fig.~\ref{fig:langevinshovarying}b) of the heuristic score shows how well distinguished the first component is from the remaining noisy components. We find that for small $\gamma \le 0.01$ and $\tau \le 0.02$, our approach successfully identifies the approximately conserved energy at intermediate time scales $T = 20$ or $T=200$ (as shown by the high rank correlation and high score prominence). For longer time scales or larger $\gamma$ and $\tau$, the approximate conservation law no longer holds and so our method identifies no conserved quantities (as shown by the low score prominence).

\section{Nonlinear Periodic Orbit of the Double Pendulum}\label{apx:doublependulumorbit}

\begin{figure}[t]
    \centering
    \includegraphics[scale=0.88]{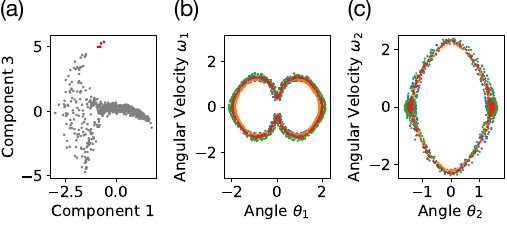}
    \caption{\textbf{Nonlinear periodic orbit of the double pendulum.} (a) The four red highlighted points in the extracted embedding correspond to  (b, c) a periodic orbit of the double pendulum that is connected to but well outside of the linear coupled oscillator regime.}
    \label{fig:doublependulumorbit}
\end{figure}

In addition to the chaotic and linear non-chaotic phases, the double pendulum can also exhibit other kinds of complex behavior, including highly nonlinear periodic orbits. In our extracted embedding (Fig.~\ref{fig:doublependulumorbit}a), we see an example of such a nonlinear periodic orbit (Figs.~\ref{fig:doublependulumorbit}b,c). The placement of this periodic orbit in the embedding also meaningfully connects it with the low energy in-phase mode from the linear coupled oscillator regime (Fig.~\ref{fig:double}g), i.e.\ this periodic orbit can be thought of as a nonlinear high energy extension of the low energy in-phase mode.

\section{Robustness to Noise \& Partial Observations}\label{apx:robust}

\begin{table}[t]
    \caption{\textbf{Rank correlations $\rho$ of linear fits with ground truth conserved quantities for the additional experiments.} $^*$The rank correlations for the low energy approximately conserved mode energies $E_\pm$ are computed on the restricted set of trajectories with first embedding component $v_1 < -1$.}
    \label{tab:robust}
    \centering
    \begin{tabular}{lll}
    \toprule
       \multirow{2}{*}{Dataset} & \multirow{2}{0.2\linewidth}{Conserved Quantity} & \multirow{2}{*}{$\rho$} \\
       &&\\
    \midrule
        \multirow{2}{0.4\linewidth}{Simple Pendulum: \emph{Position Only}} & \multirow{2}{*}{\centering $E$} & \multirow{2}{*}{\centering 0.998}\\
        &&\\[0.5em]
        \multirow{2}{0.4\linewidth}{Simple Pendulum: \emph{Position Only + Noise}} & \multirow{2}{*}{\centering $E$} & \multirow{2}{*}{\centering 0.996}\\
        &&\\[0.5em]
        \multirow{3}{0.4\linewidth}{Planar Gravitational Dynamics:\\\emph{Position Only}} & $a\cos\phi$ & 0.994\\
        & $a\sin\phi$ & 0.993\\
        & $L$ & 0.968\\[0.5em]
        \multirow{3}{0.4\linewidth}{Planar Gravitational Dynamics:\\\emph{Noise}} & $a\cos\phi$ & 0.994\\
        & $a\sin\phi$ & 0.992\\
        & $L$ & 0.945\\[0.5em]
        \multirow{3}{0.4\linewidth}{Double Pendulum:\\\emph{Position Only}} & $E$ & 0.996\\
        & $E_+$ & 0.931$^*$\\
        & $E_-$ & 0.945$^*$\\
    \bottomrule
    \end{tabular}
\end{table}

\begin{figure}[t]
    \centering
    \includegraphics[scale=0.88]{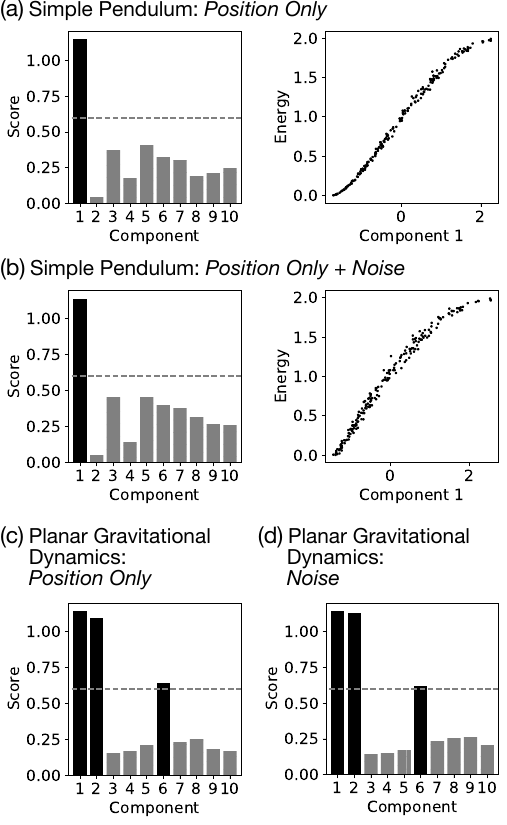}
    \caption{\textbf{Additional experiments illustrating the robustness of our approach.} (a) For the simple pendulum system, even when provided only angle $\theta$ measurement data (without angular velocity $\omega$), our method is able to identify a single relevant component corresponding to the energy the pendulum ($\rho = 0.998$). (b) If we then also add $\sigma=0.5$ Gaussian noise, we can still achieve a similar result ($\rho = 0.996$). For planar gravitational dynamics, our method also performs well given (c) only position $\mathbf{r}$ data or (d) with $\sigma=0.5$ Gaussian noise, correctly identifying the three conserved quantities.}
    \label{fig:robust}
\end{figure}

\begin{figure}[t!]
    \centering
    \includegraphics[scale=0.88]{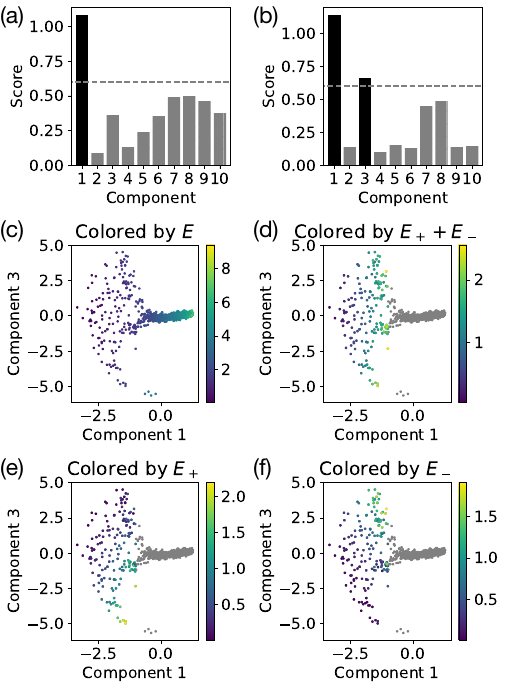}
    \caption{\textbf{Identifying conserved quantities for the double pendulum from only position data.} (a) The heuristic score (with cutoff 0.6) identifies one relevant embedding component corresponding to (b) the total energy $E$. (c) However, if we restrict the embedding to trajectories with first component $v_1 < -1$ (i.e.\ low energy trajectories) and renormalize the embedding, we find (d--f) two conserved quantities corresponding to the energies $E_\pm$ of the two decoupled low energy modes. The gray points in Figures \ref{fig:doubleposonly}d--f correspond to the high energy trajectories (first component $v_1 > -1$) which are not relevant when considering the low energy non-chaotic phase of the double pendulum.}
    \label{fig:doubleposonly}
\end{figure}

To further demonstrate the robustness of our approach, we show several additional experiments on the simple pendulum, planar gravitational dynamics, and double pendulum datasets. For the simple pendulum, our method still performs well when using only angle $\theta$ measurements, i.e.\ a partially observed phase space (Fig.~\ref{fig:robust}a). In fact, even if we add Gaussian noise (standard deviation $\sigma=0.5$) to the raw trajectory data in addition to using a partially observed phase space, we still obtain a similar result (Fig.~\ref{fig:robust}b). Similarly, for planar gravitational dynamics with only position $\mathbf{r}$ data or with added Gaussian noise ($\sigma=0.5$), our method is still able to identify the three conserved quantities (Figs.~\ref{fig:robust}c,d). For the double pendulum, we again see that we retain the same detail in the extracted embedding using only position data (Fig.~\ref{fig:doubleposonly}). The corresponding rank correlations with the ground truth conserved quantities are given in Table \ref{tab:robust}.

The robustness of our method to both noise and partial observations is largely a consequence of using the Wasserstein distance as our metric for comparing trajectories. In our problem formulation, we consider the isosurfaces with constant conserved quantities and ask for a metric for measuring distances between isosurfaces. Because the Wasserstein distance measures distances between distributions rather than disjoint isosurfaces, it can easily generalize to noisy or partially observed data where the trajectories are no longer strictly disjoint. An alternative choice of metric, e.g.\ the Hausdorff distance between sets in a metric space \cite{10.1007/978-3-030-04747-4_10}, does not have this nice property and would be much more susceptible to noise or partial observations. On the other hand, because the Wasserstein distance distinguishes between different distributions, it is susceptible to other forms of corruption such as sampling inhomogeneity (discussed in Appendix \ref{apx:samplinginhom}). In most cases, we believe that this tradeoff is worth it to retain the high degree of robustness as long as one is aware of the limitations.

\section{Comparison with Direct Fitting Methods}\label{apx:comparison}

\begin{table}[t]
    \caption{\textbf{Performance comparison of our method alongside deep learning-based direct fitting methods on the planar gravitational dynamics dataset.} We show the rank correlation $\rho$ of the embeddings with the ground truth conserved quantities. Note that the directing fitting approaches, ConservNet and Siamese Neural Network (SNN), are designed to discover a single conserved quantity and cannot handle multiple conserved quantities without dataset regeneration. Thus, with only a single dataset, we only expect the direct fitting methods to learn a single conserved quantity. For the direct fitting methods, when fitting to conserved quantities involving the orientation angle $\phi$, we choose a constant $\phi_0$ that gives the maximum $\rho$ with conserved quantities of the form $a\cos(\phi-\phi_0)$. Bolded numbers indicate the best rank correlations for each conserved quantity.}
    \label{tab:comparison}
    \centering
    \begin{tabular}{llll}
    \toprule
    \multirow{2}{*}{Dataset} & \multirow{2}{*}{Method} & \multirow{2}{0.2\linewidth}{Conserved Quantity} & \multirow{2}{*}{$\rho$} \\
       &&\\
    \midrule
    \multirow{9}{0.25\linewidth}{Clean,\\Fully Observed} & \multirow{3}{0.2\linewidth}{Manifold Learning (Ours)} & $a\cos\phi$ & \textbf{0.994}\\
        && $a\sin\phi$ & \textbf{0.992}\\
        && $L$ & \textbf{0.970}\\[0.5em]
    & \multirow{3}{0.2\linewidth}{ConservNet} & $a\cos(\phi-\phi_0)$ & 0.882\\
        && $a\sin(\phi-\phi_0)$ & 0.029\\
        && $L$ & 0.002\\[0.5em]
    &\multirow{3}{0.2\linewidth}{SNN} & $a\cos(\phi-\phi_0)$ & 0.948\\
        && $a\sin(\phi-\phi_0)$ & 0.024\\
        && $L$ & 0.007\\
    \midrule
    \multirow{9}{0.25\linewidth}{Noisy ($\sigma=0.5$), Fully Observed} & \multirow{3}{0.2\linewidth}{Manifold Learning (Ours)} & $a\cos\phi$ & \textbf{0.994}\\
        && $a\sin\phi$ & \textbf{0.992}\\
        && $L$ & \textbf{0.945}\\[0.5em]
    & \multirow{3}{0.2\linewidth}{ConservNet} & $a\cos(\phi-\phi_0)$ & 0.178\\
        && $a\sin(\phi-\phi_0)$ & 0.003\\
        && $L$ & 0.069\\[0.5em]
    &\multirow{3}{0.2\linewidth}{SNN} & $a\cos(\phi-\phi_0)$ & 0.031\\
        && $a\sin(\phi-\phi_0)$ & 0.005\\
        && $L$ & 0.025\\
    \midrule
    \multirow{9}{0.25\linewidth}{Clean, Partially Observed (Position Only)} & \multirow{3}{0.2\linewidth}{Manifold Learning (Ours)} & $a\cos\phi$ & \textbf{0.994}\\
        && $a\sin\phi$ & \textbf{0.993}\\
        && $L$ & \textbf{0.968}\\[0.5em]
    & \multirow{3}{0.2\linewidth}{ConservNet} & $a\cos(\phi-\phi_0)$ & 0.892\\
        && $a\sin(\phi-\phi_0)$ & 0.036\\
        && $L$ & 0.015\\[0.5em]
    &\multirow{3}{0.2\linewidth}{SNN} & $a\cos(\phi-\phi_0)$ & 0.916\\
        && $a\sin(\phi-\phi_0)$ & 0.010\\
        && $L$ & 0.060\\
    \bottomrule
    \end{tabular}
\end{table}

Only a few alternative approaches \cite{PhysRevLett.126.180604,PhysRevResearch.2.033499,PhysRevResearch.3.L042035} have been proposed for identifying conservation laws from trajectory samples without time information (e.g.\ any methods that require time derivative estimates of the system state are not applicable). These alternatives all fall into a broad category that we term ``direct fitting'' methods, which attempt to directly fit a parameterized function of the system state to be constant over each trajectory. This can be accomplished using a variety of methods and optimization objectives but all generally require that the system state is fully observed and that the observed trajectories have low noise. These restrictions ensure that the basic assumption of direct fitting methods---that there exists a well-defined function from the observed state to the conserved quantities---is fulfilled.

AI Poincar\'e \cite{PhysRevLett.126.180604} uses a symbolic regression approach to directly obtain an interpretable expression for the conserved quantities in terms of a library of symbolic expressions. While this method can sometimes give the exact expected conservation laws, AI Poincar\'e relies heavily on the applicability of symbolic regression, and, in cases where there is no simple symbolic expression, it will often fail. For example, for the planar gravitational dynamics dataset (named the ``Kepler problem'' in \cite{PhysRevLett.126.180604}), AI Poincar\'e fails to identify the conserved quantity associated to the orientation angle of the orbit due to its somewhat awkward symbolic representation \cite{PhysRevLett.126.180604}.

The two remaining direct fitting methods, Siamese Neural Network (SNN) \cite{PhysRevResearch.2.033499} and ConservNet \cite{PhysRevResearch.3.L042035}, both use a neural network as the parameterized function to fit the trajectory data. One important limitation of both of these approaches is their inability to discover more than a single conserved quantity per dataset. In both works \cite{PhysRevResearch.2.033499,PhysRevResearch.3.L042035}, identifying a second conserved quantity requires generating a new dataset while holding the first discovered conserved quantity fixed. Generating such a dataset as part of the method pipeline assumes both access to and fine-grained control of the data generating process, which is often not available in practice. We benchmark our manifold learning approach against these two methods on the planar gravitational dynamics dataset, variations of which (under the name ``Motion in a central potential'' and ``Kepler problem'') were also previously studied in both works \cite{PhysRevResearch.2.033499,PhysRevResearch.3.L042035}. Because we only use a single dataset, we only expect these direct fitting methods to identify a single conserved quantity. We also compare these methods on the noisy and partially observed versions of the dataset to understand their limitations.

Implementations for both direct fitting methods were adapted from code released by \cite{PhysRevResearch.3.L042035} since reference code for \cite{PhysRevResearch.2.033499} is unavailable. Following \cite{PhysRevResearch.3.L042035}, our benchmark uses a simple fully connected neural network with four hidden layers of size 320, Mish activations \cite{https://doi.org/10.48550/arxiv.1908.08681}, and an Adam optimizer \cite{kingma2015adam} with learning rate $5\times 10^{-5}$. We use a batch size of 1000 for the SNN and 200 (fixed by the number of samples per trajectory) for ConservNet. We train each network for 1000 epochs\footnote{We did not see any improvement on our dataset when the networks were trained for 50,000 epochs, as recommended by \cite{PhysRevResearch.3.L042035}.} and then compute the rank correlation $\rho$ of the fitted function with the ground truth conserved quantities (Table \ref{tab:comparison}).

The results show that, as expected, the direct fitting methods only learned a single conserved quantity for the dataset (Table \ref{tab:comparison}). Even for the single conserved quantity identified by SNN and ConservNet, we see that our manifold learning method outperforms both direct fitting methods in all settings while also identifying all three conserved quantities. Training for SNN and ConservNet took between 40--50 minutes for 1000 epochs on a single RTX 2080 Ti GPU and does not appear to benefit significantly from using larger batch sizes and additional compute resources. In comparison, our method---which is limited primarily by the Wasserstein distance estimation (Appendix \ref{apx:complexity})---takes around 40 seconds to run on eight RTX 2080 Ti GPUs and 5--6 minutes on a single RTX 2080 Ti GPU.

\section{Additional Discussion \& Limitations}\label{apx:limitations}

\subsection{Time \& Space Complexity}\label{apx:complexity}

For a dataset with $N$ trajectories and with $S$ samples per trajectory, our approach is currently limited by the computational cost of estimating the 2-Wasserstein distance for all pairs of trajectories (Table \ref{tab:timing}), giving a time complexity of $\mathcal O(N^2S^2\log S/\varepsilon^2)$ (Appendix \ref{apx:sinkhorn}). To scale to much larger datasets, we have several choices to improve the run time performance.

One simple adjustment to speed up the convergence of the Sinkhorn algorithm is to allow for a significantly larger target regularization parameter $\gamma$. The result, in fact, interpolates between the Wasserstein metric ($\gamma=0$) and a maximum mean discrepancy (MMD) metric ($\gamma=\infty$) \cite{pmlr-v89-feydy19a}. However, to improve the time complexity of our approach as we scale to large $S$, we may have to consider more approximate approaches. A popular option is to first subsample the data to form minibatches of size $s \ll S$, solve the much smaller optimal transport problem, and then average the resulting estimates for the Wasserstein distance \cite{pmlr-v84-genevay18a,pmlr-v108-fatras20a}. For a fixed number of epochs of averaging, this approach would give us linear scaling $\mathcal O(S)$ with the number of samples $S$.

To improve the scaling with the number of trajectories $N$, we may be able to take advantage of the sparse structure of the kernel matrix $K$ when the total number of conserved quantities $n \ll N$. That is, when the dimension of the embedded manifold (corresponding to the conserved quantities) is low, we expect each trajectory to have relatively few nearest neighbors, so most entries of the kernel matrix $K$ will be very close to zero for an appropriately chosen Gaussian kernel. If that is the case, we can construct the kernel matrix $K$ as a sparse matrix, e.g.\ by starting with a very coarse approximation for the pairwise Wasserstein distances and then obtaining finer estimates only for nearby trajectories or by constructing a $k$-nearest neighbor tree \cite{10.1145/355744.355745} (which takes $\mathcal O(kN\log N)$ time). With a sparse kernel matrix $K$, the diffusion map will only take $\mathcal O(N)$ time and use $\mathcal O(N)$ space.

In the future, by incorporating these approximations and algorithmic improvements, we expect to be able to adapt our approach to achieve linear scaling in time and space for very large datasets.

\begin{table}[t]
    \caption{\textbf{Run times for computing the pairwise Wasserstein distances for each dataset.} In addition to the run times, we also list the number of trajectories $N$ in the dataset, the number of samples $S$ per trajectory, and the dimension $d$ of the phase space. We used eight RTX 2080 Ti GPUs for all computations. $^\dag$Because we are interested in local conservation laws for the KdV equation dataset, the phase space is treated differently (see Sec.~\ref{sec:kdv}).}
    \label{tab:timing}
    \centering
    \begin{tabular}{lcccr}
    \toprule
    Dataset & $N$ & $S$ & $d$ & Run Time \\
    \midrule
    SHO & 200 & 200 & 2 & 11\,s\\[0.3em]
    \multirow{2}{0.23\linewidth}{Simple Pendulum} & \multirow{2}{*}{200} & \multirow{2}{*}{200} &\multirow{2}{*}{2} & \multirow{2}{*}{11\,s}\\
    &&&&\\[0.3em]
    \multirow{2}{0.23\linewidth}{Planar Grav. Dynamics} & \multirow{2}{*}{400} & \multirow{2}{*}{200} & \multirow{2}{*}{4} & \multirow{2}{*}{40\,s}\\
    &&&&\\[0.3em]
    \multirow{2}{0.23\linewidth}{Double Pendulum} & \multirow{2}{*}{1000} & \multirow{2}{*}{500} & \multirow{2}{*}{4} & \multirow{2}{*}{54\,m 43\,s}\\
    &&&&\\[0.3em]
    \multirow{2}{0.23\linewidth}{Osc. Turing Patterns} & \multirow{2}{*}{400} & \multirow{2}{*}{200} & \multirow{2}{*}{100} & \multirow{2}{*}{54\,s}\\
    &&&&\\[0.3em]
    KdV Equation & 400 & 200 & 200$^\dag$ & 26\,m 16\,s\\
    \bottomrule
    \end{tabular}
\end{table}

\subsection{Sample Complexity}\label{apx:sample}

To understand the sample complexity of our approach, we need to consider the effect of both the number of trajectories $N$ and the number of samples per trajectory $S$.

The number of samples $S$ determines the accuracy of the estimated pairwise Wasserstein distances. For the Sinkhorn algorithm, the approximation error is \cite{pmlr-v89-genevay19a}
\begin{equation}
    \mathcal O\big(S^{-1/2}\,e^{\kappa/\gamma}\big(1+\gamma^{-\lfloor d/2\rfloor}\big)\big),
\end{equation}
where $d$ is the dimension of the data, $\gamma$ is the entropy regularization parameter, and $\kappa$ is a data-dependent constant. That is, the error in our Wasserstein distance estimates scales as $1/\sqrt{S}$. Also, for $\gamma \ge 1$, notice that the dimension $d$ has no significant influence on the error bound. Furthermore, the relevant dimension $d$ in the bound is likely to be some measure of the intrinsic dimension of the data \cite{10.3150/18-BEJ1065}, allowing us to obtain good Wasserstein distance estimates even for some systems with high dimensional phase spaces (Sec.~\ref{sec:turing}).

Assuming we have accurate Wasserstein distance estimates, the spectral error of the diffusion map is $\mathcal O(N^{-2/(8+n)})$ for a manifold of dimension $n$ (i.e.\ the number of conserved quantities) \cite{doi:10.1137/20M1344093}. Thus, for integrable PDE systems like the KdV equation (Sec.~\ref{sec:kdv}) with an infinite number of conserved quantities, we must restrict ourselves to considering local conserved quantities to have a reasonable chance of reconstructing a useful embedding.

\subsection{Robustness \& Sampling Inhomogeneity}\label{apx:samplinginhom}
Because the Wasserstein distance is a metric over distributions, it has great robustness properties (Sec.~\ref{apx:robust}). However, the exact same qualities that provide this robustness also make the Wasserstein metric sensitive to sampling inhomogeneity between trajectories. That is, if two trajectories with the same conserved quantities are sampled in a way such that their distributions over the measured phase space differ, then the Wasserstein metric will treat them as different distributions, which could lead to spurious conserved quantities being identified by our method. Note that this does not include changes in the sampling process that are uniform across the trajectories and only a function of the phase space, e.g.\ sampling the state of the pendulum more often when it is near to the bottom of its swing, or sampling the position of a planet more often when it is farther from the sun for all trajectories.

One example of the relevant kind of inhomogeneity is sampling trajectories for too short a time such that the physical measure (Eq.~\ref{eq:measure}) is not well approximated. Trajectories sampled over too short a time can essentially lead our method to believe that there are additional conservation laws preventing the system state from exploring areas of phase space that may in fact be reachable for a longer trajectory. On the hand, this can be considered a feature rather than a bug in the sense that, by choosing the time over which to sample our trajectories, we are providing our method with a time scale for our conserved quantities, allowing us to probe approximately conserved quantities that appear invariant over shorter time scales (see Appendix \ref{apx:langevinsho} for an example).

If we have a fully observed phase space and low noise, then this effect can be mitigated by choosing an alternative metric, such as a Hausdorff distance \cite{10.1007/978-3-030-04747-4_10}, that does not have a strong dependence on the sampling distribution. This would essentially take advantage of the fact that any overlap between two trajectories in a fully observed phase space implies that both have the same conserved quantities. However, for noisy data or a partially observed phase space, observed overlap between measured trajectories could also be due to noise or hidden variables. Again, the same qualities that make the Hausdorff distance robust to sampling inhomogeneity also make it very sensitive to noise and partial observations.

In other words, the Wasserstein metric is able to distinguish trajectories with different conserved quantities even in a noisy partially observed phase space precisely because it can sense differences in the distributions of the trajectories.

\subsection{Physical Measures, Dissipation, \& Unbounded Dynamics}\label{apx:measure}

The assumption that the dynamical system admits a physical measure (Sec.~\ref{sec:ergodicity}) is a fairly natural one that allows us to characterize conserved quantities using invariant measures. In fact, for Hamiltonian dynamics that already admit a canonical Liouville measure \cite{goldstein2002classical}, assuming that trajectories are ergodic on the isosurfaces of conserved quantities is essentially Boltzmann's famous ergodic hypothesis \cite{pathria2021statistical,walters2000introduction}, which underpins much of statistical mechanics but is notoriously difficult to prove in general.

For dissipative systems, our assumption as stated is false since dissipation generally destroys ergodicity. However, dissipative systems still have attractors that admit physical measures \cite{medio_lines_2001,PALIS2005485}. Therefore, for a dissipative system, rather than characterizing ergodic measures on isosurfaces, our method would instead be characterizing the various attractors of the dynamics, including fixed points, limit cycles, and chaotic attractors. Our experiment with the oscillating Turing pattern (Sec.~\ref{sec:turing}) is precisely such a dissipative system with a continuous set of chaotic attractors parameterized by a spatial phase angle $\eta$.

One clear example where our assumption of a physical measure is violated in a meaningful way is when the dynamics are unbounded. For example, hyperbolic orbits following planar gravitation dynamics escape to infinity and do not converge to any invariant measure. In this case, naively applying our approach would not lead to meaningful results since, for any fixed sampling time, two different trajectories from the same hyperbolic orbit will sample very different portions of phase space depending on initial conditions. One potential solution for this issue of unbounded dynamics is to first compactify the phase space.

For example, if we compactify the traditional four dimensional Euclidean phase space of planar gravitational dynamics into a four dimensional real projective space with an attached metric (e.g.\ the Fubini-Study metric) \cite{tu2007introduction}, we have effectively made the dynamics bounded and therefore amenable to our approach. Moving to this projective representation, the physical measures on the elliptical orbits would remain largely unchanged while trajectories on hyperbolic orbits would approach a fixed point at the ``line at infinity'' (which would be a finite ``distance'' away due to the choice of metric on this compactified space). Our method would then be able to characterize the elliptical orbits by the usual three conserved quantities and hyperbolic orbits by their limiting long time behavior corresponding to their final velocity vectors as they escape to infinity.

\subsection{Trajectory Diversity \& Correlated Conserved Quantities}

One fundamental limitation, which affects not only our method but also other approaches for discovering conservation laws \cite{8618963,PhysRevLett.126.180604,PhysRevResearch.2.033499,PhysRevResearch.3.L042035,https://doi.org/10.48550/arxiv.2203.12610}, is the inability to distinguish between constraints and correlated conserved quantities. In general, constraints on the system dynamics are constant across all possible initial conditions, whereas conserved quantities vary based on initial condition. However, if there is a lack of diversity in the initial conditions for the trajectories in our dataset, then it is possible to mistake two highly correlated conserved quantities for an additional constraint on the system. For example, if the planar gravitational dynamics dataset only contained nearly circular orbits, then the angular momentum and the energy of the orbits will become highly correlated and the orientation angle will be essentially meaningless. Given such a dataset of circular trajectories and no additional information, it is not possible to distinguish between a planar gravitational dynamics dataset with poor trajectory diversity and a dataset from a system that is constrained to circular orbits with a single conserved quantity (associated with the radius of the orbits). As such, any general method for discovering conserved quantities will treat two highly correlated conserved quantities as a single conserved quantity.

\section{Dataset Details}\label{apx:dataset}

The SHO dataset contains 200 sample trajectories, each with 200 uniformly sampled states in time.

The simple pendulum dataset contains 200 trajectories with uniformly sampled energies $E \in [0, 2]$. Each trajectory has 200 sampled states at uniformly sampled times $t\in [0, 2000]$.

The planar gravitational dynamics dataset contains 400 trajectories with uniformly sampled energies $E \in [-0.15, -0.5]$, angular momenta $L \in [0, 1]$, and orbital orientation angles $\phi \in [0, 2\pi)$. Each trajectory has 200 sampled states at uniformly sampled times $t \in [0, 2000]$.

The double pendulum dataset contains 1000 trajectories with initial angles $\theta_1,\theta_2 \sim \mathrm{Unif}(-0.75\pi, 0.75\pi)$ and initial angular velocities $\omega_1,\omega_2 \sim N(0, 0.5^2)$. Each trajectory contains 500 points uniformly sampled in time $t \in [0, 50000]$. One additional subtlety of applying our approach to the double pendulum comes from the periodicity of the angles $\theta_1, \theta_2$ describing the positions of the two pendulums. The Euclidean ground metric used for optimal transport must take into account this periodicity, so we choose to leave the data unnormalized and use the shortest Euclidean distance between pairs of points in the periodic phase space.

The oscillating Turing pattern dataset contains 400 trajectories, where we initialize our states $u(x)$ and $v(x)$ with unit Gaussian noise in Fourier space and take 200 states with uniformly sampled times $t \in [300, 1300]$. By allowing for a transient time of 300, we focus our study on the long term behavior of the oscillating Turing pattern.

Finally, we study the KdV equation on a periodic domain of size $l=20$ and with mesh size $200$ (downsampled from a mesh size of $1000$ used during data generation). The dataset contains 400 trajectories each with 200 states at uniformly sampled times $t \in [0, 10]$. To produce a reasonable variety of initial conditions, each trajectory is initialized with normally distributed Fourier components scaled by a Gaussian band-limiting envelope with width uniformly sampled in the interval $[10\pi/l, 20\pi/l]$.

\onecolumn
\section{Proof of Optimal Transport for the Simple Harmonic Oscillator}\label{apx:proof}

Let the transport cost between a pair of points $(\theta_i,\theta_j) \in S^1\times S^1$ be
\begin{equation}
	c(\theta_i,\theta_j) = k_q^2(r_1\cos\theta_i - r_2\cos\theta_j)^2 + k_p^2(r_1\sin\theta_i - r_2\sin\theta_j)^2.
\end{equation}
Then, for the proposed optimal transport plan $\Pi$ with support $\Gamma$ containing all points $(\theta,\theta) \in S^1\times S^1$, we will show that $\Gamma$ is $c$-cyclically monotone, and therefore $\Pi$ is optimal. See Medio and Lines \cite{medio_lines_2001} for further details.

To demonstrate this fact, consider a finite set of pairs $\{(\theta_1, \theta_1), (\theta_2, \theta_2), \ldots,(\theta_n, \theta_n)\}\subset \Gamma$. Restricted to this finite set, the total cost given the transport plan $\Pi$ is
\begin{align}
	C &= \frac{1}{n}\sum_{i=1}^n c(\theta_i,\theta_i)\\
	&= \frac{r_1^2+r_2^2}{n}\sum_{i=1}^n(k_q^2\cos^2\theta_i + k_p^2\sin^2\theta_i) -\frac{2r_1r_2}{n}\sum_{i=1}^n (k_q^2\cos^2\theta_i +  k_p^2\sin^2\theta_i).
\end{align}

Now, consider an alternative transport plan $\Pi'$ with support $\{(\theta_1, \theta_2), (\theta_2, \theta_3), \ldots,(\theta_n, \theta_1)\}$ forming a cycle. The total cost is given by
\begin{align}
	C' &= \frac{1}{n}\sum_{i=1}^n c(\theta_i,\theta_{i+1})\\
	&= \frac{r_1^2+r_2^2}{n}\sum_{i=1}^n(k_q^2\cos^2\theta_i + k_p^2\sin^2\theta_i) -\frac{2r_1r_2}{n}\sum_{i=1}^n (k_q^2\cos\theta_i\cos\theta_{i+1} +  k_p^2\sin\theta_i\sin\theta_{i+1}),
\end{align}
where we let $\theta_{n+1} = \theta_1$. Then, the difference
\begin{align}
\begin{split}
	C' - C &= \frac{2r_1r_2k_q^2}{n}\sum_{i=1}^n \left[\frac{\cos^2\theta_i + \cos^2\theta_{i+1}}{2}  - \cos\theta_i\cos\theta_{i+1}\right]\\
	&\qquad + \frac{2r_1r_2k_p^2}{n}\sum_{i=1}^n\left[\frac{\sin^2\theta_i + \sin^2\theta_{i+1}}{2}  - \sin\theta_i\sin\theta_{i+1}\right]\\
	& \ge 0,
\end{split}
\end{align}
since
\begin{align}
	\frac{\cos^2\theta_i + \cos^2\theta_{i+1}}{2} &\ge \cos\theta_i\cos\theta_{i+1}
\intertext{and}
	\frac{\sin^2\theta_i + \sin^2\theta_{i+1}}{2} &\ge \sin\theta_i\sin\theta_{i+1}
\end{align}
by the AM--GM inequality (and is trivially true if the right hand side is negative). Therefore, any such cycle will result in an equal or higher transport cost (strictly higher if at least one pair $\theta_i,\theta_{i+1}$ are distinct), implying that $\Gamma$ is $c$-cyclically monotone.

\twocolumn

\end{appendices}


\bibliography{biblio}


\end{document}